\documentclass[aps,onecolumn,superscriptaddress,nofootinbib,notitlepage]{revtex4-1}
\usepackage{aas_macros}
\usepackage[utf8]{inputenc}
\usepackage[T1]{fontenc}
\usepackage[english,francais]{babel}
\usepackage{aas_macros}
\usepackage{amsmath}
\usepackage{amssymb}
\usepackage{graphicx}
\usepackage{bm}
\usepackage{wasysym}
\usepackage{enumitem}
\usepackage{multirow}
\usepackage[breaklinks=true,colorlinks,citecolor=blue,linkcolor=blue,urlcolor=blue]{hyperref}
\usepackage{tikz}
\usetikzlibrary[patterns]
\usetikzlibrary{shapes}

\makeatletter
\newskip\@bigflushglue \@bigflushglue = -100pt plus 1fil

\def\bigcentering{\let\\\@centercr\rightskip\@bigflushglue
\leftskip\@bigflushglue
\parindent\z@\parfillskip\z@skip}

\makeatother

\newcommand{\dd}{\mathrm{d}}

\begin{document}

\title{Determination of uncertainty profiles in neutral atmospheric properties measured by radio occultation experiments}

\author{A.~Bourgoin}
\affiliation{SYRTE, Observatoire de Paris, PSL Research University, CNRS, Sorbonne Universit\'es, UPMC P6, LNE, 61 avenue de l'Observatoire, 75014 Paris, France}
\affiliation{Département d'Astrophysique-AIM, CEA/DRF/IRFU, CNRS/INSU, Université Paris-Saclay, Université de Paris, 91191 Gif-sur-Yvette, France}
\email{adrien.bourgoin@obspm.fr}

\author{E.~Gramigna}
\affiliation{Dipartimento di Ingegneria Industriale, Alma Mater Studiorum -- Università di Bologna, Via Fontanelle 40, 47121 Forlì, Italy}

\author{M.~Zannoni}
\affiliation{Dipartimento di Ingegneria Industriale, Alma Mater Studiorum -- Università di Bologna, Via Fontanelle 40, 47121 Forlì, Italy}
\affiliation{Centro Interdipartimentale di Ricerca Industriale Aerospaziale (CIRI AERO), Alma Mater Studiorum -- Università di Bologna, Via Baldassarre Carnaccini 12, 47121 Forlì, Italy}

\author{L.~Gomez~Casajus}
\affiliation{Centro Interdipartimentale di Ricerca Industriale Aerospaziale (CIRI AERO), Alma Mater Studiorum -- Università di Bologna, Via Baldassarre Carnaccini 12, 47121 Forlì, Italy}

\author{P.~Tortora}
\affiliation{Dipartimento di Ingegneria Industriale, Alma Mater Studiorum -- Università di Bologna, Via Fontanelle 40, 47121 Forlì, Italy}
\affiliation{Centro Interdipartimentale di Ricerca Industriale Aerospaziale (CIRI AERO), Alma Mater Studiorum -- Università di Bologna, Via Baldassarre Carnaccini 12, 47121 Forlì, Italy}

\date{\today}

\begin{abstract}
  Radio occultations are commonly used to assess remotely the thermodynamic properties of planets or satellites' atmospheres within the solar system. The data processing usually involves the so-called Abel inversion method or the numerical ray-tracing technique. Both these approaches are now well established, however, they do not allow to easily determine the uncertainty profiles in the atmospheric properties, and this makes the results difficult to interpret statistically. Recently, a purely analytical approach based on the time transfer functions formalism was proposed for modeling radio occultation data. Using this formulation, we derive uncertainty relationships between the frequency shift and the thermodynamic properties of the neutral atmosphere such as the temperature, pressure, and neutral number density. These expressions are important for interpreting previous results from past radio occultation experiments. They are especially relevant for deriving the system requirements for future missions in a rigorous manner and consistently with the scientific requirements about the atmospheric properties retrieval.
\end{abstract}
 
\maketitle

\section{Introduction}
\label{sec:intro}

When an electromagnetic signal crosses an optical medium it experiences refraction. Refraction operates at two different levels within the framework of geometrical optics. First, it causes the phase of the signal to slow down or speed up while propagating in a neutral or ionized medium, respectively. Secondly, it bends the signal towards regions of higher index of refraction. Both these effects introduce delays and hence frequency shifts with respect to a signal that would have been transmitted in a vacuum.

Occultation experiments precisely exploit refraction in order to remotely retrieve the atmospheric properties of planetary atmospheres (i.e., the ionosphere and the neutral atmosphere). The basic principle is to establish a radio link between two separated elements when a planetary atmosphere stands between them. Accordingly, the electromagnetic signal is affected by the presence of the atmosphere and experiences perturbations with respect to a signal that would have been transmitted in a vacuum (i.e., in the absence of the atmosphere). By analyzing these perturbations one can retrieve the atmospheric properties of the occulting atmosphere. If the element that transmits the signal is a spacecraft radio system, the experiment is called a ``radio occultation'' (see e.g., \citet{1965Sci...149.1243K,1965JGR....70.3217F,1968P&SS...16.1035F,1985AJ.....90.1136L,1987JGR....9214987L,1992AJ....103..967L,2012Icar..221.1020S,2015RaSc...50..712S}). If the receiving element is located on a spacecraft, the radio occultation experiment is a ``spacecraft-spacecraft'' experiment (see e.g., GPS/MET experiment \citep{1999AnGeo..17..122S} or GPS/COSMIC experiment \citep{2008BAMS...89..313A}), whereas it is called a ``spacecraft-Earth'' experiment if the receiving element is located on Earth (see e.g., \citet{1971AJ.....76..123F}). In this paper, we focus on radio occultations in a spacecraft-Earth configuration.

Two methods are commonly employed for processing radio occultations data, namely the ``Abel inversion'' \citep{1968JGR....73.1819P,1971AJ.....76..123F} and the ``ray-tracing'' \citep{1981JGR....86.8721L,1992AJ....103..967L,2014JGRA..11910621G,2015RaSc...50..712S}. The Abel inversion is employed when the occulting atmosphere can be assumed to be spherically symmetric. It is an exact expression providing the index of refraction profile directly from the bending angle which is the angle between the direction of the ray path before entering the atmosphere and the direction of the ray path after exiting the atmosphere. The bending angle is itself retrieved from the frequency shift due to the presence of the atmosphere along the light path. The ray-tracing technique is an integration of the equations describing the optical rays across a layered barotropic atmosphere. The refractivity in each layer and the direction of the ray path before it entered the atmosphere (i.e., the pointing) are iteratively determined such that the computed frequency matches the observed frequency. Although the ray-tracing method is the most general one, it usually requires a very significant computational time.

Regarding statistical errors, both the Abel inversion and the ray-tracing do not allow to directly associate uncertainty profiles to the atmospheric properties that are derived. Indeed, the measurement noise is totally absorbed when the bending angle is determined or when the pointing is solved for. Therefore, the determination of uncertainties in the thermodynamic properties usually requires an \emph{ad hoc} processing. Over the past 40 years different methods have been proposed to determine the uncertainties. For instance, \citet{1979Icar...39..192L} performed a linearization of the Abel inversion model allowing them to adapt mathematical linear techniques for modeling the error propagation. This linear method is thus well suited for determining uncertainty profiles within spherically symmetric atmospheres. The method was successfully applied to past radioscience data in the context of occultations by Venus' atmosphere (see e.g.,  \citet{1994Icar..110...79J,2007Natur.450..657P,2009JGRE..114.0B36T}) or Mars' atmosphere (see e.g., \citet{1999JGR...10426997H,2016P&SS..127...44P}). Another interesting approach, which is more general than the method by \citet{1979Icar...39..192L}, is based on Monte Carlo simulations. It allows to propagate the uncertainties on input variables through any given data processing pipeline. This procedure provides more reliable statistical errors and can be applied indifferently to Abel inversion or numerical ray-tracing codes, however it usually requires a more important computational time. In the context of Abel inversion, the Monte Carlo method was successfully employed by \citet{2011Icar..215..460S,2012Icar..221.1020S} to interpret past radio occultation data of Cassini by Titan's atmosphere and by \citet{2021RaSc...5607205P} to interpret past radio occultation data of Mars Reconnaissance Orbiter (MRO) by Mars' atmosphere. However, both the linear method by \citet{1979Icar...39..192L} and the more general Monte Carlo method cannot be employed for the preliminary design of future radio occultation experiments. To overcome this issue, \citet{2010AdSpR..46...58W} derived simple relationships relating the frequency shift uncertainty to the electron density and neutral number density uncertainties. The method assumes vertical exponential profiles for both the ionosphere and the neutral atmosphere and then exploits the integral form of the Abel transform to derive the relationships. The derived uncertainty profiles are thus dependent on the scale heights of both ionosphere and neutral atmosphere. In a revised version, \citet{2020AdSpR..66.2466W} proposed an alternative approach that do not require an \emph{a priori} knowledge of the scale heights. This revised formulation, based on the Abel transform too, only depends on the vertical resolution. While in principle it could be employed to perform preliminary design of future radio occultation experiments, it still remains challenging to predict changes in the vertical resolution at the ground level without an \emph{a priori} knowledge of the scale heights. Nevertheless, these analytical approaches by Withers, even if they are less precise than purely numerical ones such as Monte Carlo simulations, have the great advantage of providing an easy, fast, and a comprehensive description of the basic relationships between the different sources of error. They are versatile and can thus be applied either to provide first order estimate of uncertainties for past radio occultations or to design future experiments.

Thus, the motivations of this work are twofold. First, with Cassini and the future JUpiter ICy Moons Explorer (JUICE) mission, the need for new methods able to process radio occultation data by an oblate and windy atmosphere is more present than ever given the improving precision of radio science observations. Up to now, only purely numerical methods can support this data processing because finding closed-form solutions for a light ray crossing a gas giants' atmosphere is a difficult task. However, an analytical solution would have many advantages with respect to a numerical one, for instance an improved computational time. Most importantly in the context of this work, the second main motivation lies in that analytical solutions are easily turned into uncertainty relationships that can directly be applied to the design of future mission unlike purely numerical methods.

Recently, it was shown by \citet{PhysRevD.101.064035}, that in the context of relativistic geometrical optics, the time and frequency transfers can be analytically determined up to any order while considering both the effects of gravity and refraction due to a neutral medium. The relativistic counterpart for ionized medium is challenging to elaborate within a covariant formalism, therefore the present paper focuses on neutral medium only. The approach by \citet{PhysRevD.101.064035} is based on two theoretical tools, namely the time transfer functions formalism \citep{2002PhRvD..66b4045L,2004CQGra..21.4463L} and the optical metric of spacetime \citep{doi101002andp19233772202}, also called Gordon's metric. The latter allows to handle refraction caused by neutral medium as spacetime curvature while the former handles theoretical problems related to the time and frequency transfers in curved spacetime. The formalism introduced by \citet{PhysRevD.101.064035} was used recently by \citet{2021A&A...648A..46B} for modeling the time and frequency transfers in radio occultation experiments. The authors derived the first order expressions of the time and frequency transfers due to a spherically symmetric atmosphere considering the relativistic light-dragging effect. This relativistic effect is naturally accounted for by the covariant formalism and represents in that sense a significant improvement with respect to previous perturbation approaches within the framework of non-relativistic geometrical optics \citep{2019A&A...624A..41B}. From the analytical solutions derived in \citet{2021A&A...648A..46B} it is possible to derive the uncertainty relationships between the frequency shift and the physical properties of the neutral atmosphere (i.e., the neutral number density, temperature, and pressure) as functions of the altitude. This is precisely the aim of the present paper where such relationships are obtained in Eqs. \eqref{eq:sigbend}--\eqref{eq:sigPTK}.

The paper is organized as follows. Section \ref{sec:assumptions} lists the notations and assumptions we make throughout the paper. Section~\ref{sec:rappels} recalls the main results of \citet{PhysRevD.101.064035} concerning the formalism of time transfer functions applied to the propagation of light across a moving neutral medium. In this section, we show that the radioscience observable quantities such as the frequency shift or the pseudo-range (i.e., the frequency and time transfers) are completely determined when the explicit form of the time transfer function is known. Section \ref{sec:application} recalls the main results of \citet{2021A&A...648A..46B} concerning the application of the time transfer functions formalism to radio occultation experiments when the occulting atmosphere is spherically symmetric. In this section, we give the explicit form of the time transfer function. The so-obtained solutions are simplified assuming an isothermal profile across the atmosphere and neglecting the relativistic light-dragging effect. In Sect. \ref{sec:uncert}, the simplified solutions are used as a starting point to derive the uncertainty relationships in atmospheric properties (cf. Eqs. \eqref{eq:sigbend}--\eqref{eq:sigPTK}). The so-obtained relationships can be applied indifferently to interpret past radio occultation experiments or to design future experiments. We put to the test our solutions by interpreting past radio occultations of Cassini by Titan's atmosphere, past occultations of Mars Global Surveyor (MGS) by Mars' atmosphere, and past occultations of Venus Express (VEX) by Venus' atmosphere. Our results are compared to analytical predictions by \citet{2010AdSpR..46...58W,2020AdSpR..66.2466W} and are also compared to outputs from more precise Monte Carlo simulations. Finally, we give our conclusions in Sect. \ref{sec:ccl}.

\section{General assumptions and notations}
\label{sec:assumptions}

This paper focuses on the propagation of light rays through a linear, isotropic, and nondispersive medium filling a spatially bounded region of spacetime $\mathcal D$. The regions of spacetime outside $\mathcal D$ are supposed to be empty of any matter. The domain $\mathcal D$ represents the limit of the neutral atmosphere.

The influence of gravity on the propagation of light is considered negligible, so $g$, the physical spacetime metric, is assumed to be a Minkowski metric. We systematically make use of an orthonormal Cartesian coordinate system $(x^\mu)=(x^0,\bm x)$ so the components of the physical metric may be written as
\begin{equation}
  g_{\mu\nu}=\eta_{\mu\nu}=\text{diag}(+1,-1,-1,-1)\text{,}
  \label{eq:flat}
\end{equation}
where Greek indices run from 0 to 3.

We set $x^0=ct$, with $c$ being the speed of light in a vacuum and $t$ a time coordinate, and we denote by $\bm x$ the triple of spatial coordinates $(x^1,x^2,x^3)$. More generally, we use the notation $\bm a=(a^1,a^2,a^3)$ for a triple constituted by the spatial components of a 4-vector and $\underbar{$\bm b$}=(b_1,b_2,b_3)$ for a triple built with the spatial components of a covariant 4-vector. According to assumption \eqref{eq:flat}, we have $\bm b=-\underbar{$\bm b$}$.

Given the triples $\bm a$, $\underbar{$\bm b$}$, and $\bm c$, the usual Euclidean scalar product $\bm a\cdot\bm c$ is denoted by $\sum_{i=1,3}a^ic^i$ and $\bm a\cdot\underbar{$\bm b$}$ denotes the quantity $\sum_{i=1,3}a^ib_i$. Furthermore, the Euclidean norm of $\bm{a}$ and $\underbar{$\bm b$}$ are denoted by $\Vert\bm a\Vert=(\bm a\cdot\bm a)^{1/2}$ and $\Vert\underbar{$\bm b$}\Vert=(\underbar{$\bm b$}\cdot\underbar{$\bm b$})^{1/2}$, respectively.

For the sake of legibility, we employ $(f)_{x}$ or $[f]_{x}$ instead of $f(x)$ whenever necessary. When a quantity $f(x)$ is evaluated at two point-events $x_A$ and $x_B$, we employ $f_{A/B}$ or $(f)_{A/B}$ to denote $f(x_A)$ and $f(x_B)$, respectively. The partial differentiations of $f$ w.r.t. $\bm x_A$ and $\bm x_B$ are denoted by $\bm\partial_{A}f$ and $\bm\partial_{B}f$, respectively. By definition, $\bm\partial_{A}$ and $\bm\partial_{B}$ are covariant triples and are always identified as such, hence, we omit the ``underbar'' for readability.

\section{Time and frequency transfers in neutral medium}
\label{sec:rappels}

In this section, the main results from \citet{PhysRevD.101.064035} are recalled and the basic relations of the present paper are inferred by substituting the metric components $g_{\mu\nu}$ from Eq. \eqref{eq:flat} into the equations provided into \citet{PhysRevD.101.064035}. Our goal is to show that the observable quantities such as the time and frequency transfers can both be expressed in term of the time transfer function. In order to prepare the discussion about radio occultation experiments, we conclude the section by applying our equations to a stationary spherically symmetric optical spacetime.

\begin{figure}
  \begin{center}
    \includegraphics[scale=0.29]{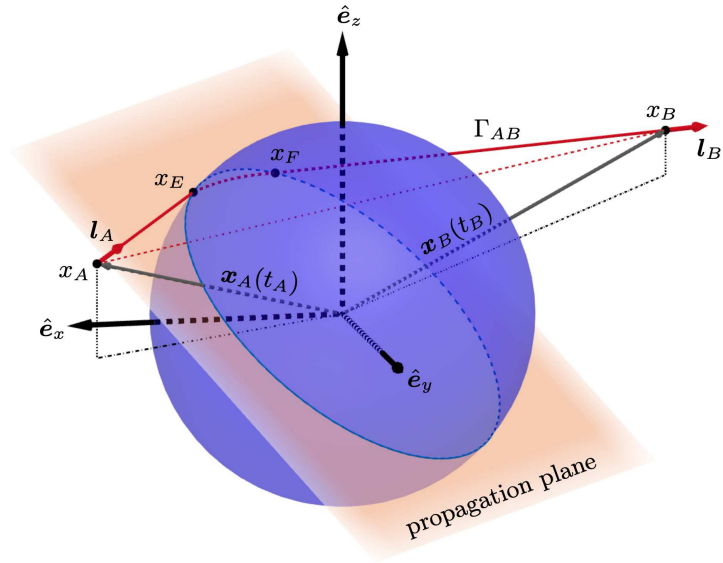}
  \end{center}
  % \begin{center}
  %   \includegraphics[scale=0.22]{occultation}
  % \end{center}
  % \setlength{\unitlength}{1.0cm}
  % \begin{picture}(0,0)
  %   \put(-4.05,3.75){\rotatebox{0}{$\hat{\bm e}_x$}}
  %   \put(0.65,3.1){\rotatebox{0}{$\hat{\bm e}_y$}}
  %   \put(-0.25,7.75){\rotatebox{0}{$\hat{\bm e}_z$}}
  %   \put(-3.8,4.5){\rotatebox{0}{$x_A$}}
  %   \put(3.7,6.55){\rotatebox{0}{$x_B$}}
  %   \put(-2.58,5.68){\rotatebox{0}{$x_E$}}
  %   \put(-1.1,6.0){\rotatebox{0}{$x_F$}}
  %   \put(-3.4,5.0){\rotatebox{0}{$\bm l_A$}}
  %   \put(4.35,6.05){\rotatebox{0}{$\bm l_B$}}
  %   \put(1.5,6.3){\rotatebox{0}{$\Gamma_{AB}$}}
  %   \put(-1.75,4.45){\rotatebox{-7}{$\bm x_A(t_A)$}}
  %   \put(0.75,4.78){\rotatebox{29.5}{$\bm x_B(t_B)$}}
  %   \put(0.6,1.2){\rotatebox{20}{propagation plane}}
  % \end{picture}
  \vspace{-0.2cm}
  \caption{Schematic illustration of a radio occultation experiment. The vectorial basis $(\hat{\bm e}_x,\hat{\bm e}_y,\hat{\bm e}_z)$ is centered at the occulting body center-of-mass and is non-rotating with respect to distant stars. The light-ray $\Gamma_{AB}$ is emitted at the point-event $x_A$ with spatial coordinates $\bm x_A(t_A)$, and is received at $x_B$ with spatial coordinates $\bm x_B(t_B)$. The point-events $x_E$ and $x_F$ are the atmosphere's entrance and exit points, respectively. The directions $\bm l_A$ and $\bm l_B$ are the unit-vectors that are tangent to $\Gamma_{AB}$ at $x_A$ and $x_B$, respectively. According to Eq. \eqref{eq:flat}, they are given by $\bm l_A=-\underbar{$\bm l$}_{A}$ and $\bm l_B=-\underbar{$\bm l$}_{B}$, respectively. For an occultation by a spherically symmetric atmosphere the propagation plane is fixed and contained the four point-events $x_A$, $x_E$, $x_F$, and $x_B$. The red dashed line joining $x_A$ and $x_B$ is the zeroth-order light-path lying on the surface of the past-light cone of $x_B$ (see also figure~1 of \citet{2021A&A...648A..46B} to appreciate the geometry of the occultation in a spacetime diagram).}
  \label{fig:occultation}
\end{figure}

\subsection{Time transfer function formalism}

Let us consider a one-way electromagnetic transfer where a light ray $\Gamma_{AB}$ is starting from an emission point-event $x_A$ of coordinates $(ct_A,\bm x_A)$ and is arriving at a reception point-event $x_B$ of coordinates $(ct_B,\bm x_B)$. We suppose that a part of $\Gamma_{AB}$ travels through the domain $\mathcal D$, while the other part travels through a vacuum, that is to say a medium such that $n=1$ with $n$ being the index of refraction. The geometry of the light pass represents an occultation experiment. It is depicted in figure~\ref{fig:occultation}.

According to \citet{2021A&A...648A..46B}, within the geometrical optics framework, the eikonal function of an electromagnetic signal passing through a neutral medium, is a first integral along $\Gamma_{AB}$ (this would no longer be the case for a ionized medium which is part of the reason why formulating a covariant theory of light propagation into a dispersive medium is challenging). This shows that $t_A$ is an implicit function of $\bm x_A$, $t_B$, and $\bm x_B$. Hence, it is appropriate to introduce $\mathcal T$, the reception time transfer function associated with $\Gamma_{AB}$, such as
\begin{equation}
  t_B-t_A=\mathcal{T}(\bm x_A,t_B,\bm x_B)\text{.}
  \label{eq:TTFdef}
\end{equation}
For the easy case of a ray of light propagating in a vacuum without gravity effects (see assumption \eqref{eq:flat}), the time transfer function is simply given by $\mathcal{T}(\bm x_A,\bm x_B)=\Vert\bm x_B-\bm x_A\Vert/c$. However, in general, when the light ray propagates through the neutral medium contained into $\mathcal D$, the time transfer function is a much more complex expression.

Let $k_\mu(x)$ be the 4-wave covector of the electromagnetic wave at $x$. An important quantity of the time transfer functions formalism is given by the 3-covector $\underbar{$\bm l$}$ which is defined from the components $k_\mu=(k_0,\underbar{$\bm k$})$. At the level of the emission and reception point-events, it is defined by
\begin{equation}
  \underbar{$\bm l$}_{A/B}=\left(\frac{\underbar{$\bm k$}}{k_0}\right)_{A/B}\text{.}
\end{equation}
Then, according to \citet{PhysRevD.101.064035}, the following useful expressions can be stated
\begin{subequations}\label{eq:kTFdef}
\begin{align}
  \underbar{$\bm l$}_{A}&=\bm\partial_{A}\left(c\mathcal{T}\right)\text{,}\label{eq:kTFA}\\
  \underbar{$\bm l$}_{B}&=-\left(1-\frac{\partial\mathcal{T}}{\partial t_B}\right)^{-1}\bm\partial_{B}\left(c\mathcal{T}\right)\text{,}\label{eq:kTFB}\\
  \frac{(k_0)_B}{(k_0)_A}&=\Bigg.1-\frac{\partial\mathcal{T}}{\partial t_B}\text{.}\label{eq:kTF0}
\end{align}
\end{subequations}
They are exact within the approximation of geometrical optics and they account for the effect of the refractive medium on the optical ray when it propagates into $\mathcal D$.

At the same approximation (i.e., the geometrical optics approximation), the frequency transfer between the emitter and the receiver is exactly given by the following relationship (see e.g., \citet{SyngeBookGR,2001A&A...370..320B}):
\begin{equation}
  \frac{\nu_B}{\nu_A}=\frac{(u^0k_0)_B}{(u^0k_0)_A}\left(\frac{1+\bm\beta_B\cdot\underbar{$\bm l$}_B}{1+\bm\beta_A\cdot\underbar{$\bm l$}_A}\right)\text{,}
  \label{eq:nu}
\end{equation}
where $\nu_{A}$ and $\nu_B$ are the emitted and received frequencies, respectively, $(u^0)_{A}$ and $(u^0)_{B}$ are the time component of $(u^\mu)_A$ and $(u^\mu)_B$, the unit 4-velocities of the emitter and receiver, respectively. The 4-velocities are by definition unit vectors for the physical metric of spacetime, hence, according to Eq.~\eqref{eq:flat}, we have the following relationship
\begin{equation}
  (u^0)_{A/B}=\left(\frac{1}{\sqrt{1-\Vert\bm{\beta}\Vert^2}}\right)_{A/B}\text{,}
  \label{eq:4vel}
\end{equation}
where $\bm\beta_{A}$ and $\bm\beta_B$ denote the coordinate 3-velocity vectors, namely
\begin{equation}
  \bm\beta_{A/B}=\left(\frac{\bm u}{u^0}\right)_{A/B}=\frac{1}{c}\left(\frac{\dd \bm x}{\dd t}\right)_{A/B}\text{.}
  \label{eq:3velAB}
\end{equation}

Interestingly, according to Eqs. \eqref{eq:nu} and \eqref{eq:kTFdef}, the frequency transfer is completely determined once the expression for the time transfer function is explicitly known (see also \citet{2002PhRvD..66b4045L,2012CQGra..29w5027H,2014PhRvD..89f4045H}). Similarly, the time transfer function completely determines the (coordinate) time transfer as seen from Eq. \eqref{eq:TTFdef}.

\subsection{Analytical expansion of the time transfer function}
\label{sec:anaexp}

In the context of atmospheric radio occultation experiments, we are mainly dealing with an optical medium with refractivity (i.e., the quantity defined such that $N=n-1$) satisfying
\begin{equation}
  N(x)\ll 1
  \label{eq:Nlow}
\end{equation}
everywhere within $x\in\mathcal D$. This assumption implies that $\Gamma_{AB}$ is almost a straight line segment joining $x_A$ to $x_B$. According to \citet{2021A&A...648A..46B}, the time transfer function can thus be decomposed as
\begin{equation}
  \mathcal T(\bm x_A,t_B,\bm x_B)=\frac{\Vert\bm x_B-\bm x_A\Vert}{c}+\frac{\Delta(\bm x_A,t_B,\bm x_B)}{c}\text{,}
  \label{eq:TTF}
\end{equation}
where $\Delta/c$ is called a delay function \citep{2008CQGra25n5020T}. In the present context, the delay function is caused by optical refraction occurring inside $\mathcal D$. The effect of the optical refraction on the radio signal is twofold as mentioned in Sect. \ref{sec:intro}. First, it changes the phase velocity of the signal creating an excess path delay, and then, it bends the signal trajectory towards regions of higher refractivity generating a geometrical delay. Other delays due to gravitational effects are neglected according to the assumption \eqref{eq:flat}, so in this paper, the delay function does not account for terms proportional to the gravitational constant $G$.

Henceforth, we introduce the parameter $N_R$ in order to keep track of the degree to which $\Gamma_{AB}$ deviates from the straight line segment. Let $N_R$ be the refractivity at a the reference point-event $x_R$, that is to say $N_R=N(x_R)$. Hereafter, it is convenient to consider that $x_R$ is the point-event where the value of the measured refractivity is the largest one (e.g., in the case of an occultation by a rocky planet, $x_R$ would be spatially close to the surface of the planet). Then, we can assume the following linear relationship
\begin{equation}
  N(x)=N_R\,\mathcal N(x)\text{,}
\end{equation}
where $\mathcal N(x)$ is a scalar function that does not depend on $N_R$ and which satisfies $0<\mathcal N(x)\leqslant 1$ for a neutral medium. It follows from Eq.~\eqref{eq:Nlow} that $N_R\ll 1$. In this case, the delay function (and hence the time transfer function) is unique \citep{PhysRevD.93.044028} and admits an analytical expansion in ascending power of $N_R$, namely
\begin{equation}
  \Delta(\bm x_A,t_B,\bm x_B)=\sum_{m=1}^{+\infty}(N_R)^m\Delta^{(m)}(\bm x_A,t_B,\bm x_B)\text{.}
  \label{eq:Del}
\end{equation}
This kind of expansion can be called a post-Minkowskian expansion similarly to what is done with gravity \citep{2008CQGra25n5020T}. In that respect, we call $N_R$ a post-Minkowskian parameter. This terminology is motivated by the so-called optical metric of spacetime whose components are proportional to $N_R$ (see Eqs. (6) of \citet{2021A&A...648A..46B} for the components of the optical metric when the physical spacetime metric reduces to Minkowski's). Hence, any deviation from the Minkowski spacetime due to refractivity is parameterized by the post-Minkowskian parameter $N_R$.

After substituting for $\Delta$ from Eq. \eqref{eq:Del} into \eqref{eq:TTF} while applying \eqref{eq:kTFdef}, we find the following expressions for the 3-covectors:
\begin{subequations}\label{eq:ldef}
\begin{align}
  \underbar{$\bm l$}_A(\bm x_A,t_B,\bm x_B)&=-\bm N_{AB}+\sum_{m=1}^{+\infty}(N_R)^m\,\underbar{$\bm l$}_A^{(m)}(\bm x_A,t_B,\bm x_B)\text{,}\\
  \underbar{$\bm l$}_B(\bm x_A,t_B,\bm x_B)&=-\bm N_{AB}+\sum_{m=1}^{+\infty}(N_R)^m\,\underbar{$\bm l$}_B^{(m)}(\bm x_A,t_B,\bm x_B)\text{,}\\
  \frac{(k_0)_B}{(k_0)_A}&=1-\sum_{m=1}^{+\infty}(N_R)^m\frac{\partial}{\partial t_B}\frac{\Delta^{(m)}(\bm x_A,t_B,\bm x_B)}{c}\text{,}\label{eq:rpk0}
\end{align}
\end{subequations}
where $\bm N_{AB}=(\bm x_B-\bm x_A)/\Vert\bm x_B-\bm x_A\Vert$ and where we introduce the post-Minkowskian terms $\underbar{$\bm l$}_A^{(m)}$ and $\underbar{$\bm l$}_B^{(m)}$ such as
\begin{subequations}\label{eq:hdef}
\begin{align}
  \underbar{$\bm l$}_A^{(m)}(\bm x_A,t_B,\bm x_B)&=\bm\partial_{A}\,\Delta^{(m)}(\bm x_A,t_B,\bm x_B)\text{,}\\
  \underbar{$\bm l$}_B^{(m)}(\bm x_A,t_B,\bm x_B)&=-\bm\partial_{B}\,\Delta^{(m)}(\bm x_A,t_B,\bm x_B)\text{.}
\end{align}
\end{subequations}

Therefore, in order to determine the time and frequency transfers up to the $m$th post-Minkowskian order, the expression of the delay function must be known at least up to the same order. The general method to derive the expression of the delay function at the desired post-Minkowskian order is presented in \citet{PhysRevD.101.064035} and is based on previous work by \citet{2008CQGra25n5020T} for the propagation of light rays in a vacuum. The method drastically simplifies when the optical spacetime is stationary.

\subsection{Stationary and spherical symmetry assumptions}
\label{sec:stationarity}

Let us consider that the coordinate system $(x^\mu)$ is attached to an inertial frame that is centered at the occulting body center-of-mass and is non-rotating with respect to distant stars (see figure \ref{fig:occultation}). In some circumstances the atmospheric properties can be considered stationary within $(x^\mu)$, and hence, the optical metric of spacetime is independent of the coordinate time. Thus, the time transfer function and the delay function become independent of $t_B$ too, so that we can write $\mathcal T(\bm x_A,\bm x_B)$ and $\Delta(\bm x_A,\bm x_B)$, respectively. In addition, the stationarity condition implies that the time component of the 4-wave covector is a first integral along $\Gamma_{AB}$ as it might be seen from Eq. \eqref{eq:rpk0}, hence $(k_0)_B-(k_0)_A=0$. Accordingly, the expression for the frequency transfer in Eq.~\eqref{eq:nu} reduces to
\begin{equation}
  \frac{\nu_B}{\nu_A}=\frac{(u^0)_B}{(u^0)_A}\left(\frac{1+\bm\beta_B\cdot\underbar{$\bm l$}_B}{1+\bm\beta_A\cdot\underbar{$\bm l$}_A}\right)\text{.}
  \label{eq:nu_sta}
\end{equation}
After substituting for $\underbar{$\bm l$}_{A}$ and $\underbar{$\bm l$}_{B}$ from Eqs. \eqref{eq:ldef} into \eqref{eq:nu_sta}, $\Delta\nu$, the frequency shift due to the atmosphere, is given by
\begin{align}
  \Delta\nu&=\left[\nu_B\right]_{\mathrm{vac}}\left\{\cfrac{1+\displaystyle\sum_{m=1}^{+\infty}(N_R)^m\cfrac{\underbar{$\bm l$}_B^{(m)}\cdot\bm\beta_B}{1-\bm\beta_B\cdot\bm N_{AB}}}{1+\displaystyle\sum_{m=1}^{+\infty}(N_R)^m\cfrac{\underbar{$\bm l$}_A^{(m)}\cdot\bm\beta_A}{1-\bm\beta_A\cdot\bm N_{AB}}}-1\right\}\text{,}
  \label{eq:Dop}
\end{align}
where $\Delta\nu=\nu_B-\left[\nu_B\right]_{\mathrm{vac}}$ with $\left[\nu_B\right]_{\mathrm{vac}}$ being the frequency that would have been received if the light ray were to propagate in a vacuum, that is to say
\begin{equation}
  \left[\nu_B\right]_{\mathrm{vac}}=\nu_A\,\frac{(u^0)_B}{(u^0)_A}\left(\frac{1-\bm\beta_B\cdot\bm N_{AB}}{1-\bm\beta_A\cdot\bm N_{AB}}\right)\text{.}
  \label{eq:DopVac1}
\end{equation}
The deviation from the frequency transfer calculated in a vacuum, namely $\Delta\nu$, is thus parameterized by terms proportional to the ascending powers of $N_R$ that represents the atmospheric contribution to the frequency shift.

If we now impose the spherical symmetry assumption, the refractivity, the delay function, and the time transfer function become radial functions. Let $K$ be the impact parameter of the zeroth-order light path with respect to the center of symmetry (i.e., the center-of-mass of the occulting planet):
\begin{equation}
  K(\bm x_A,\bm x_B)=\Vert\bm N_{AB}\times\bm x_A\Vert\text{.}
\end{equation}
In this context, Eqs. \eqref{eq:hdef} can be alternatively written as
\begin{subequations}\label{eq:hAhB}
\begin{align}
  \underbar{$\bm l$}^{(m)}_A(\bm x_A,\bm x_B)&=-\bm S_{AB}\times\bm N_{AB}\left(1+\frac{\bm N_{AB}\cdot\bm x_A}{R_{AB}}\right)\left[\frac{\partial\Delta^{(m)}}{\partial K}\right]_{(\bm x_A,\bm x_B)}\text{,}\label{eq:hA}\\
  \underbar{$\bm l$}^{(m)}_B(\bm x_A,\bm x_B)&=-\bm S_{AB}\times\bm N_{AB}\left(\frac{\bm N_{AB}\cdot\bm x_A}{R_{AB}}\right)\left[\frac{\partial\Delta^{(m)}}{\partial K}\right]_{(\bm x_A,\bm x_B)}\text{,}\label{eq:hB}
\end{align}
\end{subequations}
where we introduce $R_{AB}=\Vert\bm x_B-\bm x_A\Vert$ and where the unit-vector $\bm S_{AB}$ is defined such as
\begin{equation}
  \bm S_{AB}=-\frac{\bm N_{AB}\times\bm x_A}{\Vert\bm N_{AB}\times\bm x_A\Vert}\text{.}
  \label{eq:SAB}
\end{equation}

We now wish to determine the expression for $\Delta^{(m)}$ in the context of an occultation by a spherically symmetric atmosphere. This is the subject of the next section.

\section{Application to radio occultation experiments}
\label{sec:application}

In this section, the main results from \citet{2021A&A...648A..46B} are recalled. Our goal is to demonstrate that the time transfer function can be explicitly determined for a light ray crossing a spherically symmetric atmosphere. We focus on the first post-Minkowskian order and we consider that the optical medium is at rest in our coordinate system, in other words we neglect the relativistic light-dragging effect. We conclude the section by deriving simplified expressions for the delay function and the bending angle profiles assuming an isothermal atmosphere. These expressions can then be used as starting points to derive the uncertainty relationships.

\subsection{Modeling of the refractivity profile}
\label{sec:spheric}

Within a spherically symmetric neutral atmosphere the refractivity profile can be modeled such as
\begin{equation}
  N(r)=N_R\,\exp\left(-\frac{r-R}{H}\right)\sum_{m=0}^{d}b_mr^m\text{,}
  \label{eq:Nr}
\end{equation}
where $r=\Vert\bm x\Vert$ and $R=\Vert\bm x_R\Vert$. In this expression, $R$ is the radius of reference (i.e., the position where the measured refractivity is the largest one, cf. discussion in Sect. \ref{sec:anaexp}). We introduced the parameters $H$ and $b_m$ whose values should be determined by comparison with real observations. The polynomial coefficients $b_m$ are determined for a range of distance (e.g., for $R\leqslant r\leqslant\mathcal H$, with $\mathcal H$ the upper limit of the neutral atmosphere) and shall not be employed outside of these upper and lower limits. The degree of the polynomial expansion is denoted by $d$.

As discussed in \citet{2021A&A...648A..46B}, taking Eq.~\eqref{eq:Nr} as the input, is a reverse-order process with respect to how real radio occultation data are usually processed. Indeed, the refractivity profile is usually the output of the data processing pipeline (see e.g., Abel inversion method). Here, our approach is more closely related to a model fitting parameter method where an input refractivity modeling is first assumed and the impact on the observable quantities is determined afterwards. Proceeding this way allows us to track the occurrence of the parameters $N_R$, $H$, and $b_m$ (see Eq. \eqref{eq:Nr}) at the level of the observable quantities, namely the time delay or the bending angle. Then, if the parameters of the modeling (i.e., $N_R$, $H$, and $b_m$) can be identified to the thermodynamic properties of the atmosphere, we thus obtain direct relationships between the physical properties of the atmosphere and the observable quantities, so that the uncertainty relationships can be eventually deduced. We will see in Sect. \ref{sec:GP}, how $H$ and $b_m$ can be related to the pressure and the temperature within the atmosphere, but for now, we just consider Eq. \eqref{eq:Nr} as a convenient modeling of the refractivity profile. Indeed, for most of the known atmospheres within the solar system, the refractivity profile roughly increases exponentially with decreasing altitude which justifies the main exponential trend in Eq. \eqref{eq:Nr}. The polynomial factor allows to account for any deviation from the pure exponential trend, especially in the upper atmosphere where, in general, the neutral atmospheric scale height has a different value than in the lower atmosphere.

\citet{2021A&A...648A..46B} have shown that within the spherical symmetry assumption, the delay function can be explicitly determined at the first post-Minkowskian order. Indeed, assuming the refractivity modeling presented in Eq.~\eqref{eq:Nr} and neglecting the light-dragging effect, the first order delay function is
\begin{align}
  \Delta^{(1)}(K)&=\mathcal L_K\sum_{m=0}^{+\infty}\frac{(2m-1)!!}{2^m}\left(\frac{H}{K}\right)^{\!m}\sum_{n=0}^{m_d}Q_{m-n}B_n(K)\text{,}
  \label{eq:Del1PM} 
\end{align}
with
\begin{equation}
  \mathcal L_K=\sqrt{2\pi}\sqrt{HK}\exp\left(-\frac{K-R}{H}\right)\text{,}
  \label{eq:LK}
\end{equation}
and
\begin{equation}
  m_d=\left\{
  \begin{array}{l l}
    m & \mathrm{for}\ m< d\text{,}\\
    d & \mathrm{for}\ m\geqslant d\text{.}
  \end{array}
  \right.
\end{equation}
The non-dimensional coefficient $B_n$ is defined as
\begin{equation}
  B_n(K)=\sum_{l=n}^{d}\left(
  \begin{array}{c}
    l\\
    n
  \end{array}
  \right)b_lK^l\text{,}
\end{equation}
and the coefficient $Q_m$ is given by
\begin{equation}
  Q_m=(-1)^{m+1}\frac{(2m+1)\cdot(2m-3)!!}{2^{2m}\cdot m!}\text{.}
\end{equation}
The binomial coefficient is
\begin{equation}
  \left(
  \begin{array}{c}
    l\\
    m
  \end{array}
  \right)=\frac{l!}{m!(l-m)!}\text{,}
\end{equation}
and the double factorial \citep{garfken67math} is defined by 
\begin{subequations}
\begin{equation}
  m!!=\left\{
  \begin{array}{l l}
    m\times(m-2)\times\ldots\times 3\times 1 & \mathrm{for}\ m\ \mathrm{odd}\text{,}\\
    m\times(m-2)\times\ldots\times 4\times 2 & \mathrm{for}\ m\ \mathrm{even}\text{,}\\
    1 & \mathrm{for}\ m=-1,0\text{,}
  \end{array}
  \right.
\end{equation}
and
\begin{equation}
  (-2m-1)!!=\frac{(-1)^m}{(2m-1)!!} \qquad \mathrm{for}\ m\geqslant 1\text{.}
\end{equation}
\end{subequations}
The expression of the time transfer function is then easily obtained from Eqs. \eqref{eq:TTF} and \eqref{eq:Del}.

In the context of radio occultation experiments, $\phi$, the bending angle of the light trajectory is an important parameter that can usually be directly linked to the frequency shift. It can be defined from $\underbar{$\bm l$}_A$ and $\underbar{$\bm l$}_B$ (for small angle it is more appropriate to use the arcsine instead of the usual arccosine definition) such as
\begin{equation}
  \phi(\bm x_A,\bm x_B)=\arcsin\left[\frac{\underbar{$\bm l$}_A\times\underbar{$\bm l$}_B}{\Vert\underbar{$\bm l$}_A\Vert\,\Vert\underbar{$\bm l$}_B\Vert}\cdot\bm S_{AB}\right]\text{.}
\end{equation}
Therefore, after making use of Eqs. \eqref{eq:ldef}, it is seen that $\phi$ can be developed in ascending power of $N_R$, namely
\begin{equation}
  \phi(\bm x_A,\bm x_B)=\sum_{m=1}^{+\infty}(N_R)^m\phi^{(m)}(\bm x_A,\bm x_B)\text{.}
  \label{eq:phi1PM}
\end{equation}
At first post-Minkowskian order and within spherical symmetry assumption, we can show from Eqs. \eqref{eq:hAhB} that $\underbar{$\bm l$}_A$ and $\underbar{$\bm l$}_B$ are unit triples. Therefore, the first term of the series expansion of the bending angle is given by
\begin{equation}
  \phi^{(1)}(\bm x_A,\bm x_B)=-\left[\frac{\partial\Delta^{(1)}}{\partial K}\right]_{(\bm x_A,\bm x_B)}\text{.}
  \label{eq:phi1PMDP}
\end{equation}
After differentiating Eq.~\eqref{eq:Del1PM}, we find
%%%\begin{align}
%%%  \phi^{(1)}(\bm x_A,\bm x_B)&=\frac{\mathcal L_K}{H}\sum_{m=0}^{+\infty}\frac{(2m-1)!!}{2^m}\left(\frac{H}{K}\right)^{\!m}\sum_{n=0}^{m_d}Q_{m-n}\sum_{l=n}^d\left(
%%%  \begin{array}{c}
%%%    l\\
%%%    n
%%%  \end{array}
%%%  \right)b_lK^l\times\nonumber\\
%%%  &\times\Bigg\{1+\frac{H}{K}\left(m-l-\frac{1}{2}\right)-2\bm\Omega_K\cdot\bm S_{AB}\left[1+\frac{H}{K}\left(m-l-\frac{3}{2}\right)\right]+\mathcal O(\Omega_K^2)\Bigg\}\text{.}\label{eq:phi1}
%%%\end{align}
\begin{equation}
  \phi^{(1)}(K)=\frac{\mathcal L_K}{H}\sum_{m=0}^{+\infty}\frac{(2m-1)!!}{2^m}\left(\frac{H}{K}\right)^{\!m}\sum_{n=0}^{m_d}Q_{m-n}\sum_{l=n}^d\left(
  \begin{array}{c}
    l\\
    n
  \end{array}
  \right)b_lK^l\Bigg[1+\frac{H}{K}\left(m-l-\frac{1}{2}\right)\Bigg]\text{.}\label{eq:phi1}
\end{equation}

At first post-Minkowskian order, it is now straightforward to relate the frequency shift residuals to the bending angle. For an application within the solar system, we can always consider that the coordinate 3-velocities are small with respect to the speed of light in vacuum, that is to say $\Vert\bm\beta_A\Vert$ and $\Vert\bm\beta_B\Vert\ll 1$. Thus, at first orders in $\Vert\bm\beta_A\Vert$, $\Vert\bm\beta_B\Vert$, and $N_R$, relation \eqref{eq:Dop} simplifies to
\begin{equation}
  \frac{\Delta\nu}{\left[\nu_B\right]_{\mathrm{vac}}}=N_R\,\underbar{$\bm l$}_{B}^{(1)}\cdot\bm\beta_B-N_R\,\underbar{$\bm l$}_{A}^{(1)}\cdot\bm\beta_A\text{.}
  \label{eq:freq1PMbet}
\end{equation}
After substituting for $\underbar{$\bm l$}_{A}^{(1)}$ and $\underbar{$\bm l$}_{B}^{(1)}$ from Eqs. \eqref{eq:hAhB} into \eqref{eq:freq1PMbet} and after making use of Eqs. \eqref{eq:phi1PM} and \eqref{eq:phi1PMDP}, we eventually deduce (see also \citet{1999JGR...10426997H}, \citet{2014P&SS..101...77W}, and \citet{2020RaSc...5507046W})
\begin{equation}
  \left(\frac{\Delta\nu}{\left[\nu_B\right]_{\mathrm{vac}}}\right)_K=\phi(K)\,(\bm S_{AB}\times\bm N_{AB})\cdot\bm{\beta}_{\mathrm{eff}}\text{,}
  \label{eq:freq}
\end{equation}
where $\bm S_{AB}$ is defined in Eq. \eqref{eq:SAB} and where $\bm\beta_{\mathrm{eff}}$, the effective coordinate velocity, is given by
\begin{equation}
  \bm{\beta}_{\mathrm{eff}}=\left(1+\frac{\bm N_{AB}\cdot\bm x_A}{R_{AB}}\right)\bm\beta_A-\left(\frac{\bm N_{AB}\cdot\bm x_A}{R_{AB}}\right)\bm\beta_B\text{.}
  \label{eq:betaeff}
\end{equation}
Let us notice that when the receiver is at infinity (e.g., for one-way downlink radio occultations by Titan), the effective velocity reduces to $\lim_{r_B\to\infty}\bm{\beta}_{\mathrm{eff}}=\bm\beta_A$.

Equations \eqref{eq:freq}, \eqref{eq:phi1PM}, and \eqref{eq:phi1} describe the evolution of the frequency shift residuals as a function of the parameters of the refractivity modeling, namely $N_R$, $H$, and $b_m$. Hereafter, we intend to relate these parameters to the thermodynamic properties of the atmosphere using the ideal gas law together with the hydrostatic equilibrium approximation.

\subsection{Ideal gas law and hydrostatic equilibrium}
\label{sec:GP}

As mentioned before, it is a reasonable approximation, for most of planets or satellites within the solar system, to suppose that their refractivity profile mainly increases exponentially with decreasing altitude. This justifies having introduced the mathematical expression \eqref{eq:Nr}. The different factors in this relation can be roughly identified with the pressure and inverse of the temperature when invoking the ideal gas law:
\begin{equation}
  N(r)=N_R\,\Bigg(\frac{P}{P_R}\Bigg)_r\Bigg(\frac{T_R}{T}\Bigg)_r\text{,}
  \label{eq:idgl}
\end{equation}
where $N_R=N_vP_R/(kT_R)$ with $N_v$ the refractive volume (see e.g., \citet{1973P&SS...21.1521E}) and $k$ the Boltzmann constant. The pressure and temperature at the level $r=R$ are denoted by $P_R=P(R)$ and $T_R=T(R)$, respectively.

When assuming hydrostatic equilibrium, it is common to identify the exponential term in Eq. \eqref{eq:Nr} to the pressure profile which usually represents the largest contribution to refractivity changes across the altitude. The pressure profile can thus be roughly modeled such as (see also \cite{2014JGRA..11910621G})
\begin{equation}
  \Bigg(\frac{P}{P_R}\Bigg)_r=\exp\left(-\frac{r-R}{H}\right)\text{.}
  \label{eq:Pr}
\end{equation}
Under these conditions, $H$ behaves as the neutral atmosphere scale height since $\dd\ln(P/P_R)/\dd r=-1/H$. Hence, it might be seen after comparing Eqs. \eqref{eq:Nr} and \eqref{eq:idgl} that the inverse of the temperature profile can be identified to
\begin{equation}
  \Bigg(\frac{T_R}{T}\Bigg)_r=\sum_{m=0}^{d}b_mr^m\text{.}
  \label{eq:Tr}
\end{equation}

The interpretation of the different pieces in Eq. \eqref{eq:Nr}, in terms of pressure and temperature (such as in Eqs. \eqref{eq:Pr} and \eqref{eq:Tr}) is a strong assumption that should not be used to process real data. Indeed, it can only be valid in the lower part of the atmosphere where the neutral atmosphere scale height can be assumed constant. Outside the lower part, when the value of the scale height is changing, the real pressure profile is not a pure exponential anymore and part of it is thus absorbed into the polynomial coefficients $b_m$ which no longer describe only the inverse of the temperature profile. However, this simplified picture for the evolution of the thermodynamic quantities is a good starting point to derive uncertainty relationships since the process does not require as much precision as the real data processing. Nevertheless, let us emphasize that assumptions \eqref{eq:Pr} and \eqref{eq:Tr} are mainly valid for the lower part of the atmosphere which is precisely where we expect our uncertainty relationships to be the most accurate. Hereafter, in order to further simplify Eq. \eqref{eq:phi1}, we also assume an isothermal atmosphere.

\subsection{Isothermal assumption}
\label{sec:isoth}

Because the derivation of uncertainty profiles does not require as much precision as for the direct retrieval of atmospheric thermodynamic properties, the previous relationships can be further simplified which renders the math more tractable. Usually the pressure variation into Eq. \eqref{eq:idgl} is much more important that the changes in the temperature so that the latter can be neglected with respect to the former which varies exponentially. This simplification is equivalent to the fact of considering an isothermal profile meaning that the only polynomial coefficient $b_m$ being different from zero in Eq. \eqref{eq:Tr} is $b_0=1$. Accordingly, the refractivity profile can be simplified (see also \citet{2010AdSpR..46...58W}) such as
\begin{equation}
  N(r)=N_R\exp\left(-\frac{r-R}{H}\right)\text{,}
  \label{eq:Nisoth}
\end{equation}
In this context, the expression of the first post-Minkowskian order is simplified. After inserting $b_0=1$ and $d=0$ into Eqs.~\eqref{eq:Del1PM} and \eqref{eq:phi1} while invoking Eq.~\eqref{eq:Nisoth}, we find
\begin{equation}
  \Delta(K)=N(K)\sqrt{2\pi}\sqrt{HK}\sum_{m=0}^{+\infty}Q_{m}\frac{(2m-1)!!}{2^m}\left(\frac{H}{K}\right)^{m}+\mathcal O(N_R^2)\text{,}\label{eq:delisoth}
\end{equation}
and the expression for the bending angle becomes
\begin{equation}
  \phi(K)=N(K)\sqrt{2\pi}\sqrt{\frac{K}{H}}\sum_{m=0}^{+\infty}Q_{m}\frac{(2m-1)!!}{2^m}\left(\frac{H}{K}\right)^{m}\Bigg[1+\frac{H}{K}\left(m-\frac{1}{2}\right)\Bigg]+\mathcal O(N_R^2)\text{.}\label{eq:phiN}
\end{equation}
Equations \eqref{eq:freq} and \eqref{eq:phiN} represent the basis of the sensitivity analysis which is developed in the next section.

\section{Uncertainty profiles in atmospheric properties}
\label{sec:uncert}

In this section, we first derive the uncertainty relationships between the atmospheric properties and the frequency shift residuals. In a second step, we validate our analytical relationships with predictions from purely numerical Monte Carlo simulations computed using realistic profiles which are obtained from data processing of past radio occultation experiments. We use Cassini data for an occultation by Titan, VEX data for an occultation by Venus, and MGS data for an occultation by Mars.

\subsection{Uncertainty relationships}

From Eq. \eqref{eq:freq}, $\sigma_\phi$, the uncertainty in the bending angle, is related to $\sigma_{\Delta\nu}$, the uncertainty in the frequency shift, by
\begin{equation}
  \sigma_\phi(K)=\left(\frac{\sigma_{\Delta\nu}}{\left[\nu_B\right]_{\mathrm{vac}}}\right)_K|(\bm S_{AB}\times\bm N_{AB})\cdot\bm{\beta}_{\mathrm{eff}}|^{-1}\text{,}
  \label{eq:sigbend}
\end{equation}
where $|\cdot|$ denotes the absolute value. Let us recall that the unit vector $\bm S_{AB}$ and the effective velocity $\bm\beta_{\mathrm{eff}}$ are both defined in Eqs.~\eqref{eq:SAB} and \eqref{eq:betaeff}, respectively. From Eq. \eqref{eq:phiN}, $\sigma_N$, the uncertainty in the refractivity, is related to the uncertainty in the bending angle according to
\begin{equation}
  \sigma_N(K)=N(K)\,\left(\frac{\sigma_\phi}{\phi}\right)_K\text{.}
  \label{eq:sigN}
\end{equation}
This last equation is fundamental for deriving the relationships between the uncertainty in the frequency shift and $\sigma_T$, $\sigma_P$, and $\sigma_\kappa$, the uncertainties in temperature, pressure, and neutral number density, respectively. The uncertainty in pressure is derived from Eqs. \eqref{eq:Pr} and \eqref{eq:Nisoth}. The uncertainty in temperature can be determined from the ideal gas law in Eq.~\eqref{eq:idgl} together with the uncertainty in pressure previously derived. The uncertainty in neutral number density is derived from the neutral number density expression, namely $\kappa(K)=N(K)/N_v$. We thus obtain the following relationships:
\begin{subequations}\label{eq:sigPTK}
\begin{align}
  \sigma_P(K)&=P_R\,\frac{\sigma_N(K)}{N_R}\text{,}\label{eq:sigP}\\
  \sigma_T(K)&=T_R\left[1+\frac{T(K)}{T_R}\right]\left(\frac{\sigma_N}{N
  }\right)_K\text{,}\label{eq:sigT}\\
  \sigma_\kappa(K)&=\frac{\sigma_N(K)}{N_v}\text{.}
\end{align}
\end{subequations}
Equation \eqref{eq:sigP} assumes isothermal assumption and is, in that sense, expressed at zeroth-order in the temperature profile (more precisely, it is expressed at zeroth order in $T_R/T-1$). Instead, to derive Eq. \eqref{eq:sigT} we had to consider the first order in the temperature profile in the ideal gas law relation.

While interpreting results from past radio occultation experiments, $\phi(K)$, $N(K)$, $P(K)$, and $T(K)$ should be taken from atmospheric profiles that are determined from data analysis using the Abel inversion method or the ray-tracing technique. However, if the aim is to perform preliminary design studies of future radio occultation experiments, the atmospheric profiles can be taken from the theoretical modeling presented in Sect.~\ref{sec:isoth}. Then, $N(K)$ and $\phi(K)$ can be modeled according to Eqs.~\eqref{eq:Nisoth} and \eqref{eq:phiN}, respectively, assuming an \emph{a priori} knowledge of the neutral atmospheric scale height (see also \citet{2010AdSpR..46...58W}). The temperature and pressure profiles can be modeled assuming the isothermal hypothesis which leads to $T(K)/T_R=1$ and $P(K)/P_R=N(K)/N_R$, respectively. Hereafter, we compare our analytical expressions with predictions from Monte Carlo simulations using past radio occultation experiments, therefore $\phi(K)$, $N(K)$, $P(K)$, and $T(K)$ are obtained from the real data processing using the Abel inversion method.

\subsection{Application to past radio occultation experiments}

We now wish to validate our analytical expressions for uncertainties in atmospheric properties by studying results of representative radio occultation experiments performed at Titan, Mars, and Venus by the Cassini, MGS, and VEX missions, respectively. We consider the four following data set. Firstly, the ingress radio occultation of Cassini by Titan on June 22nd, 2009, recorded by the Deep Space Station (DSS)-14 in one-way X-band downlink. Secondly, the ingress radio occultation of MGS by Mars on December 27th 1998, recorded by the DSS-25 in one-way X-band downlink. Finally, two egress radio occultations of VEX by Venus on August 22nd, 2006 recorded by the DSA-1 and February 1st, 2014 recorded by the DSS-45, both in one-way X-band downlink. VEX 2006 data are used to validate our own Abel inversion software with respect to the results presented in the literature, while VEX 2014 data are used for testing our analytical uncertainties. Cassini and MGS data are available on NASA's Planetary Data System (PDS) (\url{https://pds.nasa.gov/}). VEX 2006 data are available on ESA's Planetary Science Archive (PSA) (\url{https://www.cosmos.esa.int/web/psa/psa-introduction}), while VEX 2014 data were provided to us by the Multi-mission "Planetary Radar and Radio Sciences Group" at Jet Propulsion Laboratory (JPL), California  Institute  of  Technology.

\begin{figure*}
 \includegraphics[scale=0.8]{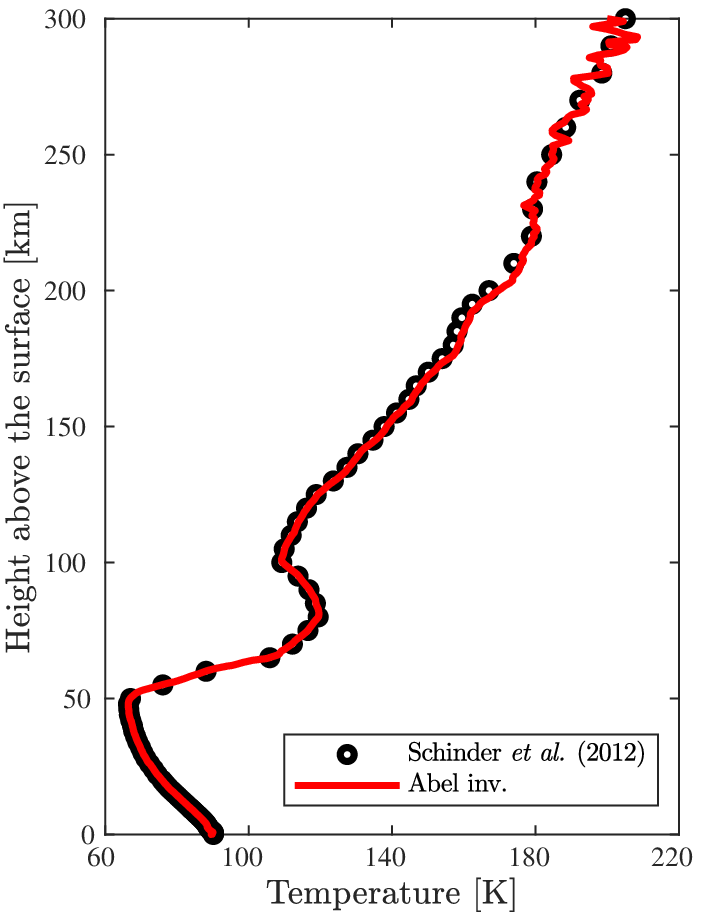}\hspace{0.25cm}
 \includegraphics[scale=0.8]{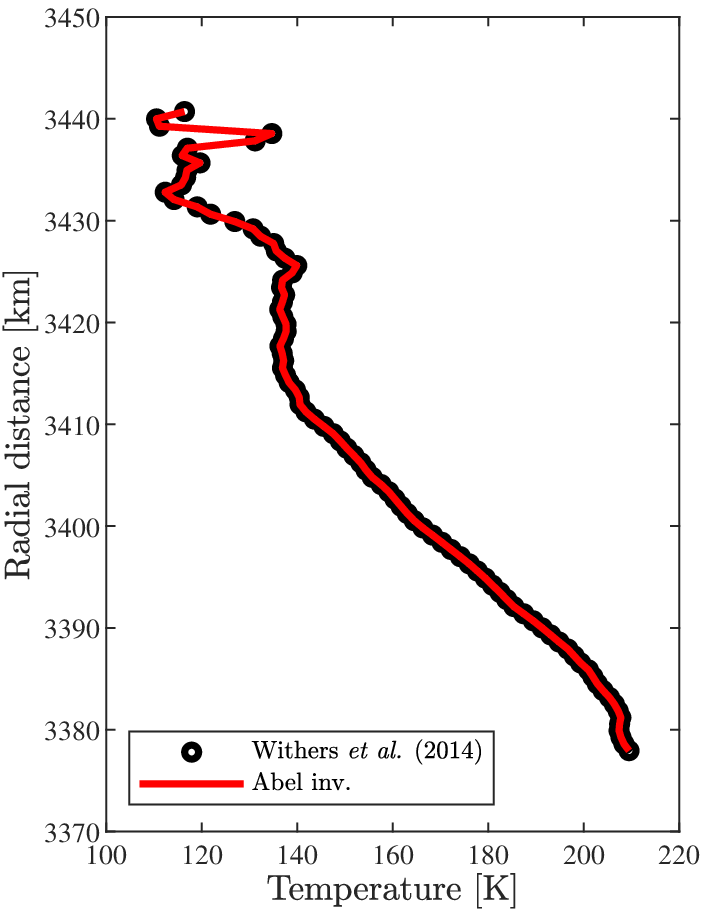}\hspace{0.25cm}
 \includegraphics[scale=0.8]{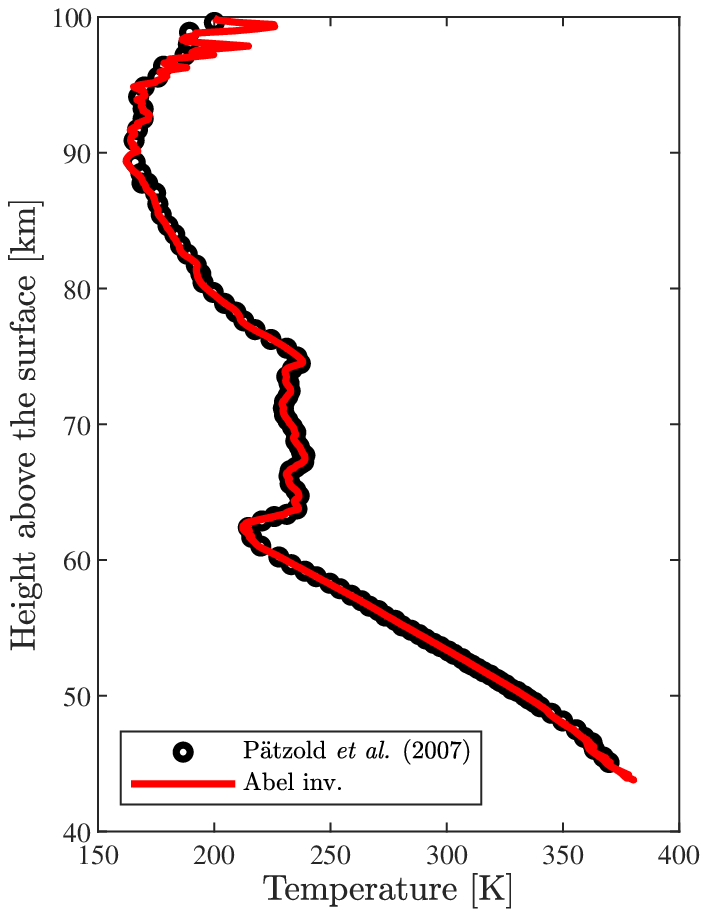}
 \centering
 \caption{Temperature profiles in the atmospheres of Titan (\emph{left panel}), Mars (\emph{middle panel}) and Venus (\emph{right panel}). The radioscience data correspond to an occultation of Cassini on June 22nd, 2009, an occultation of MGS on December 27, 1998, and an occultation of VEX on August 22, 2006. The red curve is determined from our own Abel inversion software. The black dots represent the temperature profile published in \citet{2012Icar..221.1020S} (cf. \emph{left panel} for Titan), \citet{2014P&SS..101...77W} (cf. \emph{middle panel} for Mars) and \citet{2007Natur.450..657P} (cf. \emph{right panel} for Venus) for the corresponding occultations. For these particular occultations at Titan, Mars, and Venus, the radii of reference are $R=2\,575\ \mathrm{km}$, $R=3\,377.95\ \mathrm{km}$, and $R=6\,051.8\ \mathrm{km}$, respectively.}
 \label{fig:schinder}
\end{figure*}

Owing to the fact that the three atmospheres are stationary and spherically symmetric to a good extent, the refractivity is derived by invoking the Abel inversion method. The computation of the thermodynamic profiles is performed using our own Abel inversion software. The frequency shift is modeled from the ephemerides of the spacecraft and planets, and these are given by NASA's SPICE toolkit Navigation and Ancillary Information Facility (NAIF) \citep{1996P&SS...44...65Aq}. We mention in passing the alternative approach by \citet{2021RaSc...5607205P}, where the radio occultation processing is performed directly into a precise orbit determination software which provides the spacecraft state vector trajectory. This approach allows for a more accurate determination of the spacecraft state vector during the processing of the radio occultation data.

\begin{figure*}
  \includegraphics[scale=0.9]{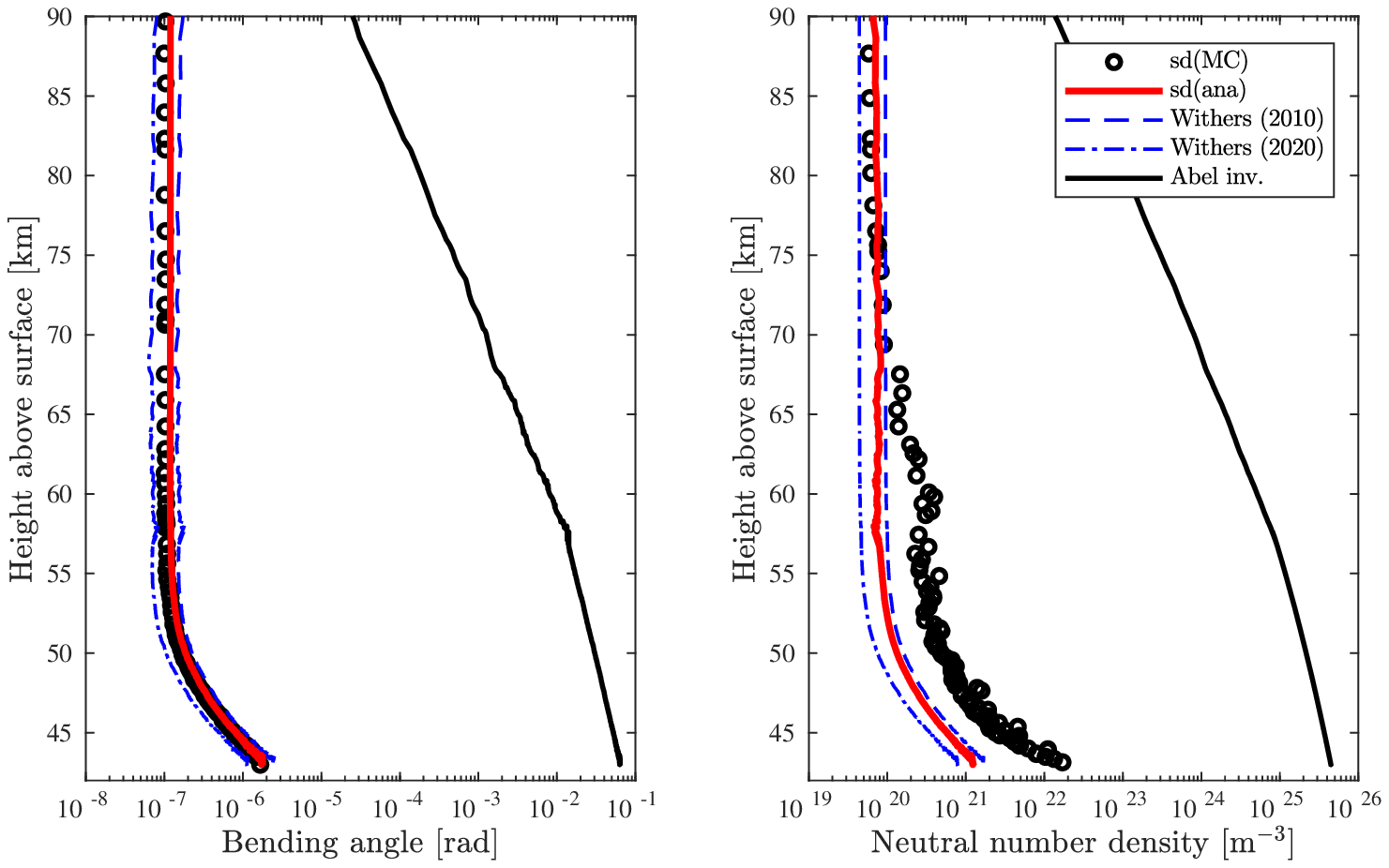}\vspace{1cm}
  \includegraphics[scale=0.9]{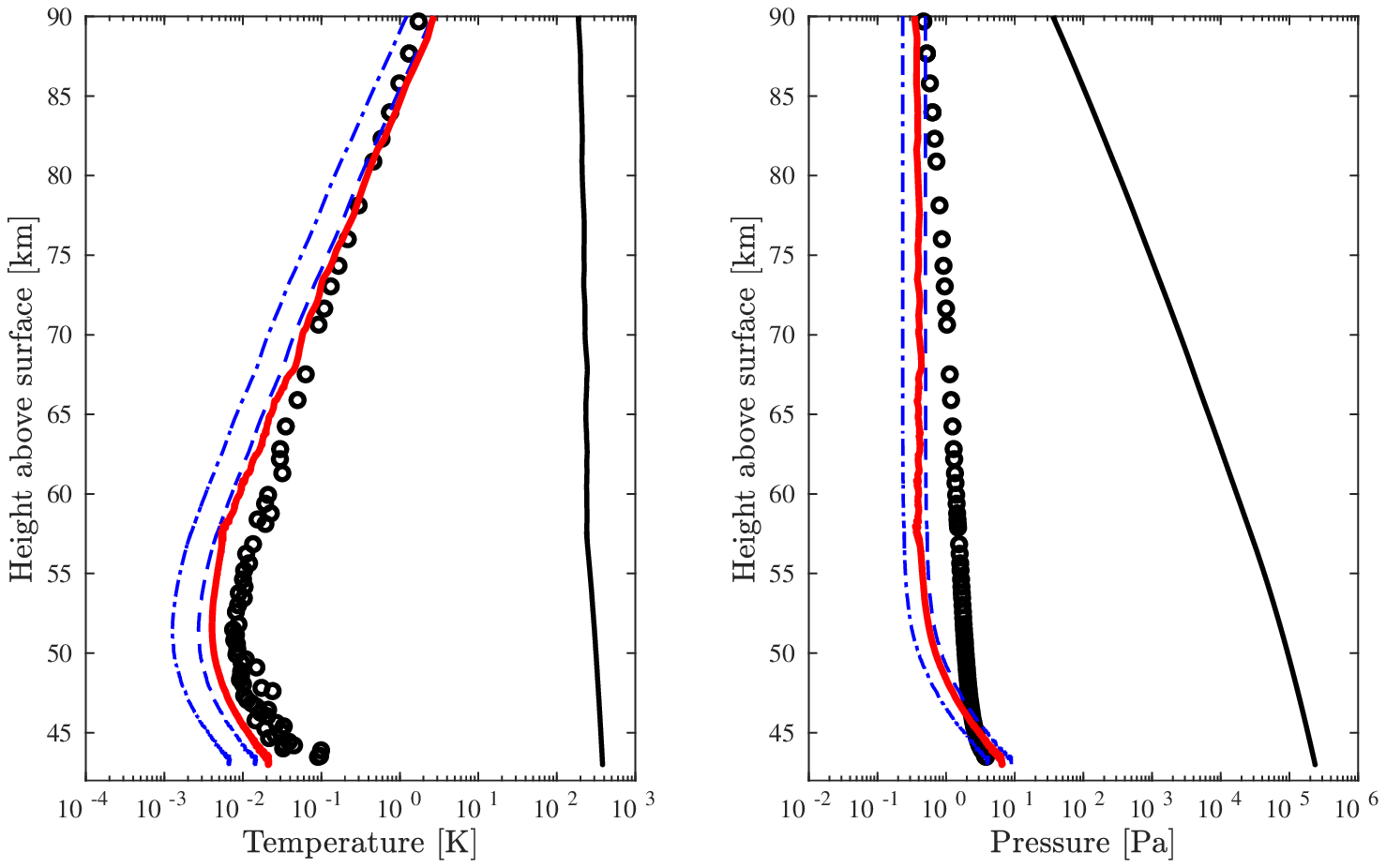}  
  \centering
  \caption{Uncertainties in bending angle (\emph{upper left panel}), neutral number density (\emph{upper right panel}), temperature (\emph{lower left panel}), and pressure (\emph{lower right panel}) for the occultation of VEX by Venus' atmosphere that occurred on February 1st, 2014. The plain black curve represents the physical profile obtained from the Abel inversion method. The black dots represent the Monte Carlo reference uncertainties, the red curve shows our analytic isothermal uncertainty obtained from Eqs. \eqref{eq:sigbend} and \eqref{eq:sigPTK} and the blue dashed line and the dashed-dotted line represent the curves obtained from \citet{2010AdSpR..46...58W} and \citet{2020AdSpR..66.2466W}, respectively. For this particular occultation at Venus, the radius of reference is $R=6\,051.8\ \mathrm{km}$.}
  \label{errorbendvex}
\end{figure*}

\begin{figure*}
  \includegraphics[scale=0.9]{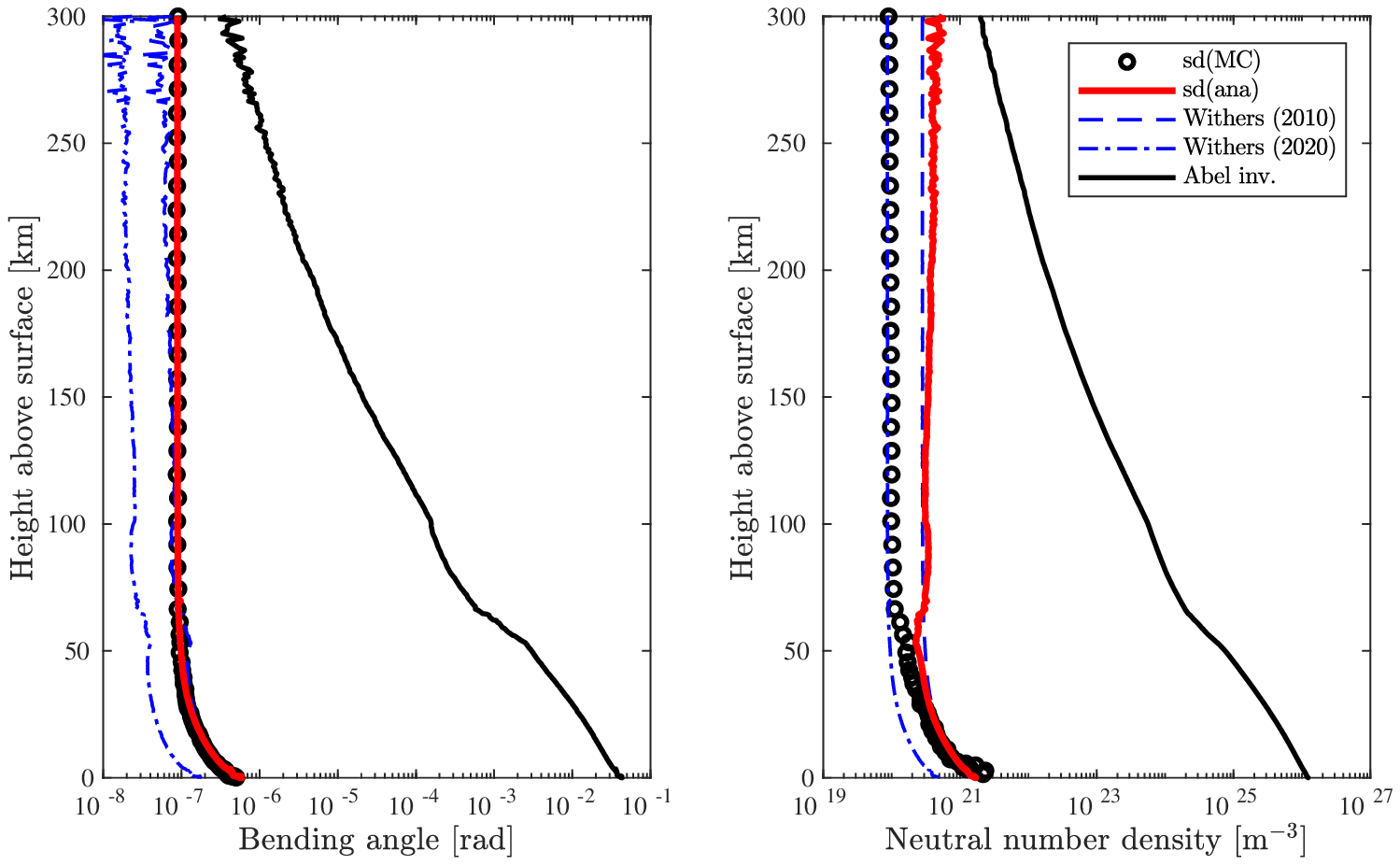}\vspace{1cm}
  \includegraphics[scale=0.9]{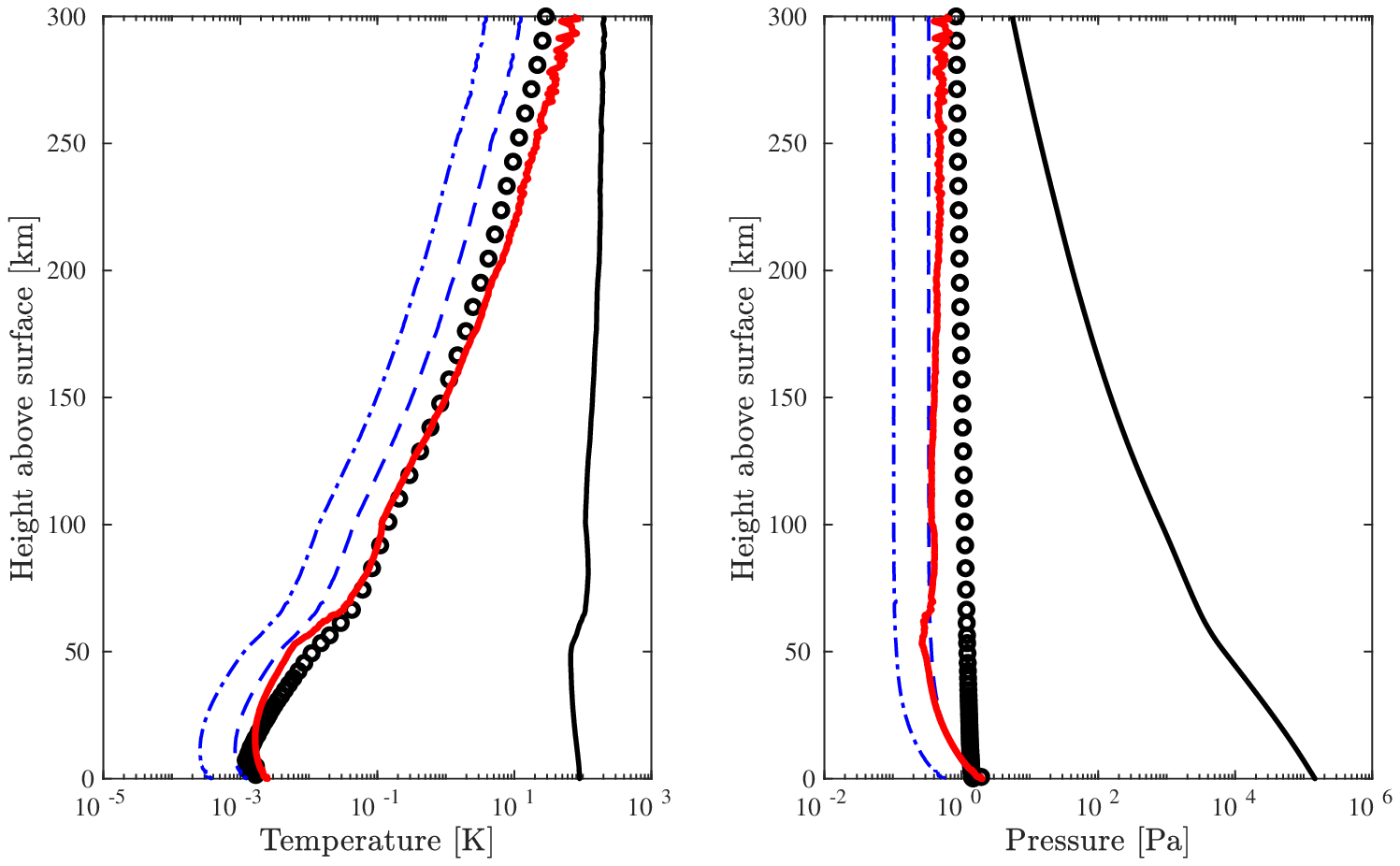}
  \centering
  \caption{Uncertainties in the bending angle (\emph{upper left panel}), neutral number density (\emph{upper right panel}), temperature (\emph{lower left panel}), and pressure (\emph{lower right panel}) for the Cassini-Titan occultation. See caption of figure \ref{errorbendvex} for additional details. For this particular occultation at Titan, the radius of reference is $R=2\,575\ \mathrm{km}$.}
  \label{errorbend}
\end{figure*}

\begin{figure*}
  \includegraphics[scale=0.9]{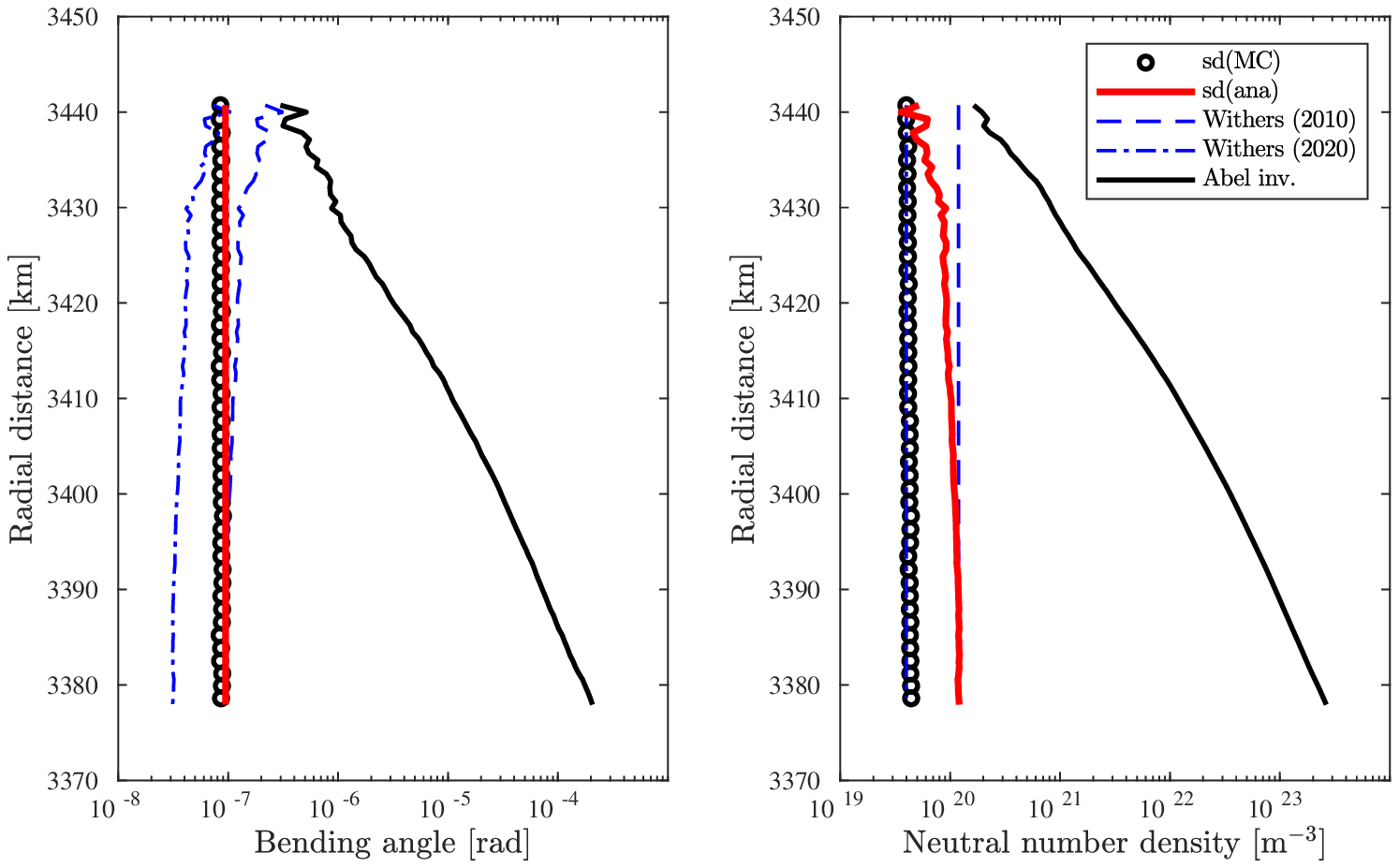}\vspace{1cm}
  \includegraphics[scale=0.9]{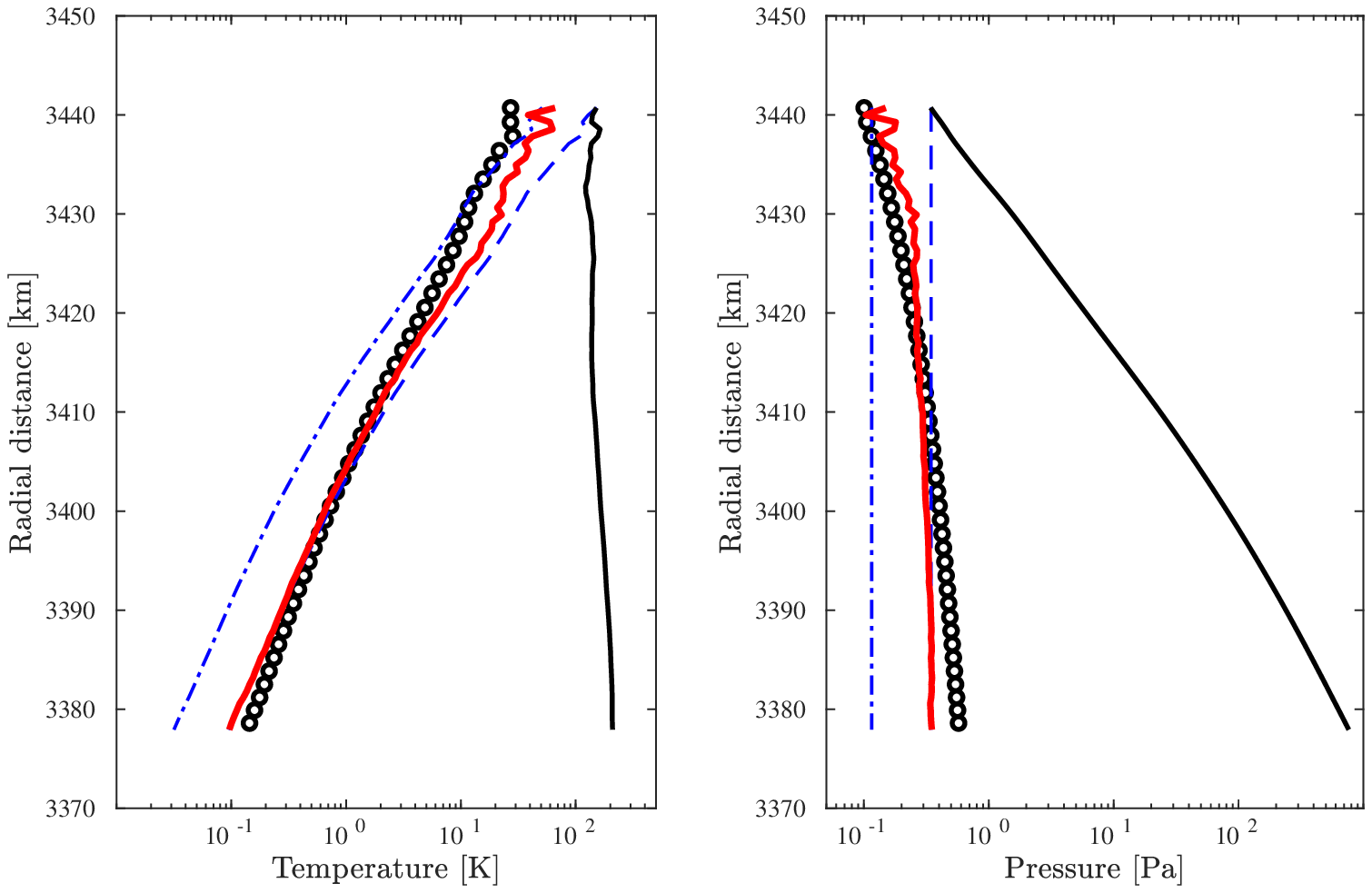}  
  \centering
  \caption{Uncertainties in the bending angle (\emph{upper left panel}), neutral number density (\emph{upper right panel}), temperature (\emph{lower left panel}), and pressure (\emph{lower right panel}) for the MGS-Mars occultation. See caption of figure \ref{errorbend} for additional details. For this particular occultation at Mars, the radius of reference is $R=3\,377.95\ \mathrm{km}$.}
  \label{errorbendmgs}
\end{figure*}

Then, for the cases of Mars and Venus, the thermodynamic properties (i.e., the neutral number density, the pressure, and the temperature) can be inferred by making use of hydrostatic equilibrium assumption together with the ideal gas law. For the case of Titan, the thermodynamic properties are derived assuming the hydrostatic equilibrium assumption and an equation of state that accounts for non-ideal gas effects in Titan's lower atmosphere \citep{1973Cryo...13...11B,1981Cryo...21..157B}, similarly to what is done in \citet{2012Icar..221.1020S}. These profiles have been validated by comparing them to those published in the literature (see e.g., \citet{2011Icar..215..460S,2012Icar..221.1020S} for Titan, \citet{1968P&SS...16.1035F,2014P&SS..101...77W,1999JGR...10426997H} for Mars, and \citet{1971AJ.....76..123F,2007Natur.450..657P,2021arXiv210208300G} for Venus). In figure \ref{fig:schinder}, we show the comparisons for the temperature profiles within the atmospheres of Titan, Mars, and Venus. The atmospheric profile for Titan is derived up to its surface so that $R$, the altitude of reference, is the equatorial radius. For Mars, the profile is computed as a function of the radial distance since the equatorial and polar radii differ by tens of kilometers. For Venus, the profiles cannot be computed below $45\ \mathrm{km}$ altitude due to low value of the signal-to-noise ratio. Therefore, the altitude of reference in this case is $45\ \mathrm{km}$ above the surface. The profiles $\phi(K)$, $N(K)$, $P(K)$, and $T(K)$ that are needed for evaluating our analytical solutions in Eqs. \eqref{eq:sigbend}, \eqref{eq:sigN}, and \eqref{eq:sigPTK} are determined from our own Abel inversion software and are depicted as plain black curves labeled ``Abel inv.'' in figures \ref{errorbendvex}, \ref{errorbend}, and \ref{errorbendmgs} for Venus, Titan, and Mars, respectively.

%%%\begin{figure}
%%%  \includegraphics[scale=0.8]{ErreurTempPresvex}
%%%  \centering
%%%  \caption{Uncertainties in temperature (left) and pressure (right) for the Venus Express-Venus occultation}
%%%  \label{errortempvex}
%%%\end{figure}

%%%\begin{figure}
%%%  \includegraphics[scale=0.8]{ErreurTempPres}
%%%  \centering
%%%  \caption{Uncertainties in temperature (left) and pressure (right) for the Cassini-Titan occultation.}
%%%  \label{errortemp}
%%%\end{figure}

%%%\begin{figure}
%%%  \includegraphics[scale=0.8]{ErreurTempPresmgs}
%%%  \centering
%%%  \caption{Uncertainties in temperature (left) and pressure (right) for the Mars Global Surveyor-Mars occultation}
%%%  \label{errortempmgs}
%%%\end{figure}

Hereafter, we compare our analytical uncertainties to reference uncertainties that are obtained from a Monte Carlo simulations \citep{2011Icar..215..460S,2012Icar..221.1020S}. The Monte Carlo method allows to propagate the uncertainties in the frequency shift through the data processing pipeline. The methodology to derive Monte Carlo estimates proceeds as follows. First, a random white Gaussian noise, $\sigma_{\Delta\nu}(K)$, is added to the original time-series of the frequency residuals, $\Delta\nu(K)$. The expression for the standard deviation of the Gaussian noise is
\begin{equation}
  \sigma_{\Delta\nu}(K)=\mathrm{N}_{\mathrm{dist}}(0,1)\sqrt{\sigma_{\mathrm{phase}}^2+\sigma_{\mathrm{thermal}}^2(K)}\text{,}
\end{equation}
where $\mathrm{N}_{\mathrm{dist}}(0,1)$ refers to a randomly selected number among the normal distribution with zero mean and unit variance. The two variances under the square root represent the contributions from the phase noise and the thermal noise. The standard deviation of the phase noise, $\sigma_{\mathrm{phase}}$, is mainly constant across the atmospheric profile and is given by the Allan deviation of the frequency residuals outside the occulting atmosphere (i.e., the baseline). The thermal noise, $\sigma_{\mathrm{thermal}}(K)$, evolves as the reciprocal of the square root of the signal-to-noise  ratio. Because the latter varies across the atmospheric profile, the thermal noise is not constant in general. We refer to \citet{2010AdSpR..46...58W} for additional comments about the different sources of noise and for an expression (cf. Eq. (23)) of the thermal noise in term of the signal-to-noise ratio.

Then, 2\,000 (adding more profiles does not significantly change the results) independent determinations of atmospheric properties are performed using the Abel inversion method, each time for a different realization of the random noise. In addition, the boundary temperature at the top of the neutral atmosphere is randomly chosen with a $\pm 10\ \mathrm{K}$ uncertainty range since the computation of the temperature and pressure profile requires an \emph{a priori} guess of the initial value of the temperature (see e.g., discussion in \citet{1973P&SS...21.1521E,1999JGR...10426997H,2003P&SS...51..541W,2020RaSc...5507046W,2021RaSc...5607205P}). Finally, the standard deviations of the bending angle, the neutral number density, the temperature, and the pressure profiles are assessed. These uncertainty profiles of reference are depicted as black dots labeled ``sd(MC)'' in figures~\ref{errorbendvex}, \ref{errorbend}, and \ref{errorbendmgs}, for Venus, Titan, and Mars, respectively. Let us emphasize that only the phase noise contribution is considered for the MGS occultation by Mars' atmosphere since no information is available on the signal-to-noise ratio for the observation we are considering.

By invoking Eqs. \eqref{eq:sigbend}--\eqref{eq:sigPTK}, we determine the predicted uncertainties in atmospheric properties by making use of the profiles $\phi(K)$, $N(K)$, $P(K)$, and $T(K)$ determined with our Abel inversion code. These analytical estimates are depicted as plain red curves labeled ``sd(ana)'' in figures \ref{errorbendvex}, \ref{errorbend}, and \ref{errorbendmgs}, for Venus, Titan, and Mars, respectively. Our results are also compared to the uncertainty predictions by \citet{2010AdSpR..46...58W,2020AdSpR..66.2466W}. Results by Withers are depicted as blue dashed curves and blue dash-dotted curves labeled ``Withers (2010)'' and ``Withers (2020)'', respectively. Let us emphasize that Withers only provides $\sigma_\kappa$, the neutral number density uncertainty so we used relations \eqref{eq:sigbend}--\eqref{eq:sigPTK} in order to infer Withers' uncertainty predictions for the bending angle, pressure, and temperature profiles, even if these quantities were not provided in the original Withers' papers.

From a general point of view, our analytical predictions in Eqs. \eqref{eq:sigbend}, \eqref{eq:sigN}, and \eqref{eq:sigPTK} show really good agreement with respect to the reference uncertainty profiles inferred from the Monte Carlo simulations. The largest differences are observed for the dense atmosphere of Venus (cf. figure \ref{errorbendvex}). For instance, around $45\ \mathrm{km}$ altitude the predicted analytical uncertainty is $\pm 5.7 \times10^{20}\ \mathrm{m^{-3}}$ for the neutral number density whereas the Monte Carlo simulations return a value of $\pm 2.4 \times10^{21}\ \mathrm{m^{-3}}$ (we report in table \ref{tab:error}, the profile values at surface for Titan, at radial distance of $3\,377.95\ \mathrm{km}$ for Mars, and at $45\ \mathrm{km}$ altitude for Venus). Therefore, for dense neutral atmospheres such as Venus, the first order analytical expressions provide slightly optimistic uncertainties, up to a factor $4.2$ in the neutral number density profile (cf. table \ref{tab:error}). For the temperature, the analytical uncertainty is optimistic by a factor $2.2$, and the most plausible explanation is that the analytical estimates were determined at first order in $N_R$ (cf. Sect. \ref{sec:application}) while second order terms might become important to account for when dealing with low altitudes within dense atmospheres. We note that both analytical methods proposed by Withers cannot exactly replicate the Monte Carlo uncertainty profiles in Venus' atmosphere, too. The reason is probably due to the fact that the solutions were derived assuming low refractivity. Let us emphasize that our analytical uncertainty in the bending angle is in really good agreement with Monte Carlo simulations, even for the dense atmosphere of Venus. The difference remains at the level one part in $10^2$. This is due to the fact that the bending is directly related to the frequency shift residuals (according to Eq. \eqref{eq:freq}) and is independent of the isothermal assumption.

For the cases of Titan and Mars, the agreement between the analytical uncertainty profiles and Monte Carlo's is excellent (cf. figures \ref{errorbend} and \ref{errorbendmgs}, respectively). At Titan's surface, the least accurate prediction is the uncertainty in the neutral number density profile which is only optimistic by a factor of 1.6. At Mars' radial distance of $3\,377.95\ \mathrm{km}$, the least accurate prediction is the uncertainty in the neutral number density which is conservative up to a factor of 2.7.

\begin{table}
  \centering
  \caption{Table of errors for the bending, the neutral number density, the temperature, and the pressure. Values for Titan are given at the surface, for Mars at a radial distance of $3\,377.95\ \mathrm{km}$, while for Venus are given at $45\ \mathrm{km}$ altitude.}
  \begin{tabular}{c|c||ccc}
    \hline\hline
    \multicolumn{2}{c||}{Profiles} & Titan & Mars & Venus \\
    \hline
    \multirow{3}{*}{$\phi\ [\mathrm{rad}]$} & Abel & $4.4\times10^{-2}$ & $2.0\times 10^{-4}$ & $5.2\times10^{-2}$ \\
%    \cline{2-5}
    & $\mathrm{sd}(\mathrm{ana})$ & $5.5\times10^{-7}$ & $9.5\times10^{-8}$ & $7.7\times10^{-7}$ \\
%    \cline{2-5}
    & $\mathrm{sd}(\mathrm{MC})$ & $5.0\times10^{-7}$ & $8.6\times10^{-8}$ & $7.4\times10^{-7}$\\
    \hline
    \multirow{3}{*}{$\kappa\ [\mathrm{m}^{-3}]$} & Abel & $1.2\times 10^{26}$ & $2.6\times10^{23}$ & $3.7\times 10^{25}$ \\
%    \cline{2-5}
    & $\mathrm{sd}(\mathrm{ana})$ & $1.4\times10^{21}$ & $1.2\times10^{20}$ & $5.7\times10^{20}$ \\
%    \cline{2-5}
    & $\mathrm{sd}(\mathrm{MC})$ & $2.2\times10^{21}$ & $4.4\times10^{19}$ & $2.4\times10^{21}$\\
    \hline
    \multirow{3}{*}{$T\ [\mathrm{K}]$} & Abel & $89.6$ & $209.6$ & $363.3$ \\
%    \cline{2-5}
    & $\mathrm{sd}(\mathrm{ana})$ & $2.1\times10^{-3}$ & $1.0\times 10^{-1}$ & $1.2\times 10^{-2}$  \\
%    \cline{2-5}
    & $\mathrm{sd}(\mathrm{MC})$ & $1.7\times10^{-3}$ & $1.4\times 10^{-1}$ & $2.6\times10^{-2}$ \\
    \hline
    \multirow{3}{*}{$P\ [\mathrm{Pa}]$} & Abel & $1.5\times10^{5}$ & $763.2$ & $1.8\times10^5$ \\
%    \cline{2-5}
    & $\mathrm{sd}(\mathrm{ana})$ & $1.9$ & $3.5\times10^{-1}$ & $3.0$ \\
%    \cline{2-5}
    & $\mathrm{sd}(\mathrm{MC})$ & $1.5$ & $5.7\times10^{-1}$ & $2.7$ \\
    \hline\hline    
  \end{tabular}
  \label{tab:error}
\end{table}

For the three cases presented here (Venus, Titan, and Mars), let us emphasize that the analytical estimates always succeed in reproducing the general trend of the Monte Carlo uncertainties. This is also true for Venus' temperature and pressure profiles, even though the analytical predictions are derived from first order expressions in $N_R$, $\Vert\bm\beta_A\Vert$, and $\Vert\bm\beta_B\Vert$. Finally, let us mention that our analytical estimates are consistent, most of the time, with the ones derived by \citet{2010AdSpR..46...58W}. The origin of this behavior becomes clear when relation~\eqref{eq:phiN} is compared to Eq.~\,(35) of \citet{2010AdSpR..46...58W}. The difference between the two expressions, that were derived using two different frameworks, is in the order of $H/K$ which is small in general. However, our analytical uncertainties, especially in temperature and pressure, remain closer to the Monte Carlo estimates than Withers' predictions. This is particularly visible for cases of Titan and Mars. However, it must be emphasized that the method by \citet{2020AdSpR..66.2466W} seems to be in really good agreement with the Monte Carlo predictions for the neutral number density profile when the thermal noise is neglected, as shown in figure~\ref{errorbendmgs}.

\section{Conclusions}
\label{sec:ccl}

In this work, we present new relationships for finding the uncertainties in neutral atmosphere properties measured by radio occultation experiments. Our approach is based on results of previous works (see \citet{PhysRevD.101.064035,2021A&A...648A..46B}) that aimed at determining the time and frequency transfers to a high degree of precision when the electromagnetic signal is crossing a neutral medium. By considering the first order solutions of \citet{2021A&A...648A..46B}, we are able to derive analytical expressions describing the evolution of the delay function and the bending angle during an occultation by a stationary and spherically symmetric atmosphere. The delay function and the bending angle are needed to determine explicit expressions for the time and frequency transfers. Noticing that, in general, the refractivity profile is dominated by pressure variations before temperature changes across the profile, we simplify the relationships assuming an isothermal atmosphere. We thus provide simplified equations (cf. Eqs.~\eqref{eq:delisoth} and \eqref{eq:phiN}) that are then used as starting points for deriving uncertainty relationships. The uncertainty relationships allow to relate the errors in the frequency shift to those in the bending angle, the refractivity, the neutral number density, the temperature, and the pressure. The relationships are given explicitly in Eqs. \eqref{eq:sigbend}, \eqref{eq:sigN}, and \eqref{eq:sigPTK}. We validate these from results of past radio occultation experiments performed at Titan, Mars, and Venus by the Cassini, MGS, and VEX missions, respectively. We thus perform Monte Carlo simulations in order to assess precise numerical uncertainty estimates that are then compared to our analytical predictions in Eqs. \eqref{eq:sigbend}, \eqref{eq:sigN}, and \eqref{eq:sigPTK}. The analytical predictions prove to replicate very accurately the numerical results from the Monte Carlo simulations. For the case of dense atmospheres, such as Venus, although the analytical error profiles can be optimistic at the lowest altitude for the pressure and neutral number density, they nevertheless capture well the trend of the uncertainty profiles (see figure~\ref{errorbendvex}). Concerning the case of more tenuous atmospheres, such as the ones of Titan and Mars, the analytical uncertainties are in really good agreement with Monte Carlo simulations (see figures~\ref{errorbend} and \ref{errorbendmgs}). We also compared our analytical uncertainty relationships with other two analytical approaches developed by \citet{2010AdSpR..46...58W} and \citet{2020AdSpR..66.2466W}. These methods provide uncertainty for the neutral number density only, so we used our relationships to extrapolate Withers' results for the temperature, pressure and bending angle. For neutral number density and pressure, our analytical predictions are similar to those by \citet{2010AdSpR..46...58W} especially at low altitudes. Indeed, even though our approach and Withers' approach rely on completely different formalisms (\citet{2010AdSpR..46...58W} is based on Abel inversion integrals whereas this work uses the time transfer functions formalism applied to optical spacetimes) the relationships that are obtained are eventually similar when the spherical symmetry assumption and the low refractivity approximation are applied. The other method by \citet{2020AdSpR..66.2466W} seems to agree the best with the Monte Carlo predictions for the neutral number density in Mars' atmosphere where we neglected the thermal noise contribution. Our method, by providing uncertainties also for temperature and pressure, is thus complementary to \citet{2010AdSpR..46...58W,2020AdSpR..66.2466W}. Withers' methods consider the characteristics of the experiment outside the atmosphere to retrieve uncertainties. However, our method allows to obtain higher accuracy uncertainty profiles since it depends on the characteristics of the experiment and its variations along the whole atmosphere. In addition, the complete experiment occultation geometry is taken into account as a function of the altitude. As a matter of fact, our analytical prediction for the uncertainty in the bending angle is in really good agreement with the Monte Carlo predictions, even for the more complex Venus case.

As demonstrated in this work the uncertainty relationships derived in Eqs. \eqref{eq:sigbend}, \eqref{eq:sigN}, and \eqref{eq:sigPTK}, can be used for interpreting results of past radio occultation experiments. In this attempt, the different atmospheric property profiles $\phi(K)$, $N(K)$, $P(K)$, and $T(K)$, that are needed for the evaluation, can be directly taken from the data analysis which is performed with the Abel inversion method or the ray-tracing technique. Therefore, the application of Eqs.~\eqref{eq:sigbend}, \eqref{eq:sigN}, and \eqref{eq:sigPTK} does not require to implement an \emph{ad hoc} pipeline dedicated to the computation of uncertainties since the thermodynamic profiles are already available from the data processing. The advantages of our uncertainty relationships with respect to purely numerical approaches such as \cite{1979Icar...39..192L} or Monte Carlo simulations, is thus twofold since they can also be employed for performing preliminary design studies of future radio occultation experiments. In that case, the profiles can be derived from theoretical modelings presented in Sect.~\ref{sec:isoth}; as for \citet{2010AdSpR..46...58W}, an \emph{a priori} knowledge of the scale height is needed, though.

Finally, let us emphasize the fact that our method is purely analytical and is obviously less precise than a purely numerical approach. Our final relationships are analytical first order expressions in $N_R$, $\Vert\bm\beta_A\Vert$, and $\Vert\bm\beta_B\Vert$, in addition, we have assumed the perfect gas law and we have supposed an isothermal profile to make the math more tractable. However, because we did each assumption at the right moment, we are still able to reproduce fully numerical results to a really good extent for Mars and Titan. The case of Venus is obviously more difficult to get; since the atmosphere of Venus is dense, the first approximation in $N_R$ or the isothermal assumption make our results slightly optimistic at low altitude. However, even with all the simplifications, our final expressions (cf. Eqs. \eqref{eq:sigbend}, \eqref{eq:sigN}, and \eqref{eq:sigPTK}) are able to reproduce results from complex numerical simulations and offer an easy and a comprehensive alternative to the numerical methods. In addition, they can also be applied to the design of future occultation experiments.

To conclude, our method could be further improved by releasing the isothermal assumption and considering the full solutions of Sect. \ref{sec:spheric}. In addition, a more complete solution could be considered including higher order terms in $N_R$ which would allow to predict more reliably the uncertainty profiles in denser atmospheres such as Venus'. The formalism could also be applied to non-spherical symmetry and by considering the relativistic light-dragging effect, which was already derived in \citet{2021A&A...648A..46B}, the obtained uncertainty relationships could be applied to occultations by giant gas planets, too. This would be of particular interest for the future data processing of the JUICE mission and to the design of future missions aiming at exploring the outer solar system.

\section{Acknowledgments}
The authors are grateful to the Italian Space Agency (ASI) for financial support through Agreement No. 2018-25-HH.O in the context of ESA's JUICE mission, and Agreement No. 2020-13-HH.O in the context of TRIDENT's mission Phase A study. A.B. is grateful to Centre National d'\'Etudes Spatiales (CNES) for financial support. E.G. is grateful to “Fondazione Cassa dei Risparmi di Forlì” for financial support of his PhD fellowship. The authors are grateful to Dustin Buccino and Kamal Oudrhiri from the Multi-mission "Planetary Radar and Radio Sciences Group" at Jet Propulsion Laboratory (JPL), California Institute of Technology, for providing the Venus Express radio occultation data used in this paper. The authors are also thankful to Andrea Caruso for providing simulated data for testing the Monte-Carlo simulations. Finally, the  authors acknowledge the crucial contribution of the radio science teams of the Cassini, MGS and VEX missions, for planning the radio occultation experiments described in this paper, acquiring the data, publishing their results and then archiving the data in public repositories. Without their hard work and dedication, this paper would never have seen the light.

\bibliographystyle{apsrev4-1}
\bibliography{error_budget}

\begin{thebibliography}{45}%
\makeatletter
\providecommand \@ifxundefined [1]{%
 \@ifx{#1\undefined}
}%
\providecommand \@ifnum [1]{%
 \ifnum #1\expandafter \@firstoftwo
 \else \expandafter \@secondoftwo
 \fi
}%
\providecommand \@ifx [1]{%
 \ifx #1\expandafter \@firstoftwo
 \else \expandafter \@secondoftwo
 \fi
}%
\providecommand \natexlab [1]{#1}%
\providecommand \enquote  [1]{``#1''}%
\providecommand \bibnamefont  [1]{#1}%
\providecommand \bibfnamefont [1]{#1}%
\providecommand \citenamefont [1]{#1}%
\providecommand \href@noop [0]{\@secondoftwo}%
\providecommand \href [0]{\begingroup \@sanitize@url \@href}%
\providecommand \@href[1]{\@@startlink{#1}\@@href}%
\providecommand \@@href[1]{\endgroup#1\@@endlink}%
\providecommand \@sanitize@url [0]{\catcode `\\12\catcode `\$12\catcode
  `\&12\catcode `\#12\catcode `\^12\catcode `\_12\catcode `\%12\relax}%
\providecommand \@@startlink[1]{}%
\providecommand \@@endlink[0]{}%
\providecommand \url  [0]{\begingroup\@sanitize@url \@url }%
\providecommand \@url [1]{\endgroup\@href {#1}{\urlprefix }}%
\providecommand \urlprefix  [0]{URL }%
\providecommand \Eprint [0]{\href }%
\providecommand \doibase [0]{http://dx.doi.org/}%
\providecommand \selectlanguage [0]{\@gobble}%
\providecommand \bibinfo  [0]{\@secondoftwo}%
\providecommand \bibfield  [0]{\@secondoftwo}%
\providecommand \translation [1]{[#1]}%
\providecommand \BibitemOpen [0]{}%
\providecommand \bibitemStop [0]{}%
\providecommand \bibitemNoStop [0]{.\EOS\space}%
\providecommand \EOS [0]{\spacefactor3000\relax}%
\providecommand \BibitemShut  [1]{\csname bibitem#1\endcsname}%
\let\auto@bib@innerbib\@empty
%</preamble>
\bibitem [{\citenamefont {{Kliore}}\ \emph {et~al.}(1965)\citenamefont
  {{Kliore}}, \citenamefont {{Cain}}, \citenamefont {{Levy}}, \citenamefont
  {{Eshleman}}, \citenamefont {{Fjeldbo}},\ and\ \citenamefont
  {{Drake}}}]{1965Sci...149.1243K}%
  \BibitemOpen
  \bibfield  {author} {\bibinfo {author} {\bibfnamefont {A.}~\bibnamefont
  {{Kliore}}}, \bibinfo {author} {\bibfnamefont {D.~L.}\ \bibnamefont
  {{Cain}}}, \bibinfo {author} {\bibfnamefont {G.~S.}\ \bibnamefont {{Levy}}},
  \bibinfo {author} {\bibfnamefont {V.~R.}\ \bibnamefont {{Eshleman}}},
  \bibinfo {author} {\bibfnamefont {G.}~\bibnamefont {{Fjeldbo}}}, \ and\
  \bibinfo {author} {\bibfnamefont {F.~D.}\ \bibnamefont {{Drake}}},\ }\href
  {\doibase 10.1126/science.149.3689.1243} {\bibfield  {journal} {\bibinfo
  {journal} {Science}\ }\textbf {\bibinfo {volume} {149}},\ \bibinfo {pages}
  {1243} (\bibinfo {year} {1965})}\BibitemShut {NoStop}%
\bibitem [{\citenamefont {{Fjeldbo}}\ and\ \citenamefont
  {{Eshleman}}(1965)}]{1965JGR....70.3217F}%
  \BibitemOpen
  \bibfield  {author} {\bibinfo {author} {\bibfnamefont {G.}~\bibnamefont
  {{Fjeldbo}}}\ and\ \bibinfo {author} {\bibfnamefont {V.~R.}\ \bibnamefont
  {{Eshleman}}},\ }\href {\doibase 10.1029/JZ070i013p03217} {\bibfield
  {journal} {\bibinfo  {journal} {\jgr}\ }\textbf {\bibinfo {volume} {70}},\
  \bibinfo {pages} {3217} (\bibinfo {year} {1965})}\BibitemShut {NoStop}%
\bibitem [{\citenamefont {{Fjeldbo}}\ and\ \citenamefont
  {{Eshleman}}(1968)}]{1968P&SS...16.1035F}%
  \BibitemOpen
  \bibfield  {author} {\bibinfo {author} {\bibfnamefont {G.}~\bibnamefont
  {{Fjeldbo}}}\ and\ \bibinfo {author} {\bibfnamefont {V.~R.}\ \bibnamefont
  {{Eshleman}}},\ }\href {\doibase 10.1016/0032-0633(68)90020-2} {\bibfield
  {journal} {\bibinfo  {journal} {\planss}\ }\textbf {\bibinfo {volume} {16}},\
  \bibinfo {pages} {1035} (\bibinfo {year} {1968})}\BibitemShut {NoStop}%
\bibitem [{\citenamefont {{Lindal}}\ \emph {et~al.}(1985)\citenamefont
  {{Lindal}}, \citenamefont {{Sweetnam}},\ and\ \citenamefont
  {{Eshleman}}}]{1985AJ.....90.1136L}%
  \BibitemOpen
  \bibfield  {author} {\bibinfo {author} {\bibfnamefont {G.~F.}\ \bibnamefont
  {{Lindal}}}, \bibinfo {author} {\bibfnamefont {D.~N.}\ \bibnamefont
  {{Sweetnam}}}, \ and\ \bibinfo {author} {\bibfnamefont {V.~R.}\ \bibnamefont
  {{Eshleman}}},\ }\href {\doibase 10.1086/113820} {\bibfield  {journal}
  {\bibinfo  {journal} {\aj}\ }\textbf {\bibinfo {volume} {90}},\ \bibinfo
  {pages} {1136} (\bibinfo {year} {1985})}\BibitemShut {NoStop}%
\bibitem [{\citenamefont {{Lindal}}\ \emph {et~al.}(1987)\citenamefont
  {{Lindal}}, \citenamefont {{Lyons}}, \citenamefont {{Sweetnam}},
  \citenamefont {{Eshleman}},\ and\ \citenamefont
  {{Hinson}}}]{1987JGR....9214987L}%
  \BibitemOpen
  \bibfield  {author} {\bibinfo {author} {\bibfnamefont {G.~F.}\ \bibnamefont
  {{Lindal}}}, \bibinfo {author} {\bibfnamefont {J.~R.}\ \bibnamefont
  {{Lyons}}}, \bibinfo {author} {\bibfnamefont {D.~N.}\ \bibnamefont
  {{Sweetnam}}}, \bibinfo {author} {\bibfnamefont {V.~R.}\ \bibnamefont
  {{Eshleman}}}, \ and\ \bibinfo {author} {\bibfnamefont {D.~P.}\ \bibnamefont
  {{Hinson}}},\ }\href {\doibase 10.1029/JA092iA13p14987} {\bibfield  {journal}
  {\bibinfo  {journal} {\jgr}\ }\textbf {\bibinfo {volume} {92}},\ \bibinfo
  {pages} {14987} (\bibinfo {year} {1987})}\BibitemShut {NoStop}%
\bibitem [{\citenamefont {{Lindal}}(1992)}]{1992AJ....103..967L}%
  \BibitemOpen
  \bibfield  {author} {\bibinfo {author} {\bibfnamefont {G.~F.}\ \bibnamefont
  {{Lindal}}},\ }\href {\doibase 10.1086/116119} {\bibfield  {journal}
  {\bibinfo  {journal} {\aj}\ }\textbf {\bibinfo {volume} {103}},\ \bibinfo
  {pages} {967} (\bibinfo {year} {1992})}\BibitemShut {NoStop}%
\bibitem [{\citenamefont {{Schinder}}\ \emph {et~al.}(2012)\citenamefont
  {{Schinder}}, \citenamefont {{Flasar}}, \citenamefont {{Marouf}},
  \citenamefont {{French}}, \citenamefont {{McGhee}}, \citenamefont {{Kliore}},
  \citenamefont {{Rappaport}}, \citenamefont {{Barbinis}}, \citenamefont
  {{Fleischman}},\ and\ \citenamefont {{Anabtawi}}}]{2012Icar..221.1020S}%
  \BibitemOpen
  \bibfield  {author} {\bibinfo {author} {\bibfnamefont {P.~J.}\ \bibnamefont
  {{Schinder}}}, \bibinfo {author} {\bibfnamefont {F.~M.}\ \bibnamefont
  {{Flasar}}}, \bibinfo {author} {\bibfnamefont {E.~A.}\ \bibnamefont
  {{Marouf}}}, \bibinfo {author} {\bibfnamefont {R.~G.}\ \bibnamefont
  {{French}}}, \bibinfo {author} {\bibfnamefont {C.~A.}\ \bibnamefont
  {{McGhee}}}, \bibinfo {author} {\bibfnamefont {A.~J.}\ \bibnamefont
  {{Kliore}}}, \bibinfo {author} {\bibfnamefont {N.~J.}\ \bibnamefont
  {{Rappaport}}}, \bibinfo {author} {\bibfnamefont {E.}~\bibnamefont
  {{Barbinis}}}, \bibinfo {author} {\bibfnamefont {D.}~\bibnamefont
  {{Fleischman}}}, \ and\ \bibinfo {author} {\bibfnamefont {A.}~\bibnamefont
  {{Anabtawi}}},\ }\href {\doibase 10.1016/j.icarus.2012.10.021} {\bibfield
  {journal} {\bibinfo  {journal} {\icarus}\ }\textbf {\bibinfo {volume}
  {221}},\ \bibinfo {pages} {1020} (\bibinfo {year} {2012})}\BibitemShut
  {NoStop}%
\bibitem [{\citenamefont {{Schinder}}\ \emph {et~al.}(2015)\citenamefont
  {{Schinder}}, \citenamefont {{Flasar}}, \citenamefont {{Marouf}},
  \citenamefont {{French}}, \citenamefont {{Anabtawi}}, \citenamefont
  {{Barbinis}},\ and\ \citenamefont {{Kliore}}}]{2015RaSc...50..712S}%
  \BibitemOpen
  \bibfield  {author} {\bibinfo {author} {\bibfnamefont {P.~J.}\ \bibnamefont
  {{Schinder}}}, \bibinfo {author} {\bibfnamefont {F.~M.}\ \bibnamefont
  {{Flasar}}}, \bibinfo {author} {\bibfnamefont {E.~A.}\ \bibnamefont
  {{Marouf}}}, \bibinfo {author} {\bibfnamefont {R.~G.}\ \bibnamefont
  {{French}}}, \bibinfo {author} {\bibfnamefont {A.}~\bibnamefont
  {{Anabtawi}}}, \bibinfo {author} {\bibfnamefont {E.}~\bibnamefont
  {{Barbinis}}}, \ and\ \bibinfo {author} {\bibfnamefont {A.~J.}\ \bibnamefont
  {{Kliore}}},\ }\href {\doibase 10.1002/2015RS005690} {\bibfield  {journal}
  {\bibinfo  {journal} {Radio Science}\ }\textbf {\bibinfo {volume} {50}},\
  \bibinfo {pages} {712} (\bibinfo {year} {2015})}\BibitemShut {NoStop}%
\bibitem [{\citenamefont {{Steiner}}\ \emph {et~al.}(1999)\citenamefont
  {{Steiner}}, \citenamefont {{Kirchengast}},\ and\ \citenamefont
  {{Ladreiter}}}]{1999AnGeo..17..122S}%
  \BibitemOpen
  \bibfield  {author} {\bibinfo {author} {\bibfnamefont {A.~K.}\ \bibnamefont
  {{Steiner}}}, \bibinfo {author} {\bibfnamefont {G.}~\bibnamefont
  {{Kirchengast}}}, \ and\ \bibinfo {author} {\bibfnamefont {H.~P.}\
  \bibnamefont {{Ladreiter}}},\ }\href {\doibase 10.1007/s00585-999-0122-5}
  {\bibfield  {journal} {\bibinfo  {journal} {Annales Geophysicae}\ }\textbf
  {\bibinfo {volume} {17}},\ \bibinfo {pages} {122} (\bibinfo {year}
  {1999})}\BibitemShut {NoStop}%
\bibitem [{\citenamefont {{Anthes}}\ \emph {et~al.}(2008)\citenamefont
  {{Anthes}}, \citenamefont {{Bernhardt}}, \citenamefont {{Chen}},
  \citenamefont {{Cucurull}}, \citenamefont {{Dymond}}, \citenamefont
  {{Ector}}, \citenamefont {{Healy}}, \citenamefont {{Ho}}, \citenamefont
  {{Hunt}}, \citenamefont {{Kuo}}, \citenamefont {{Liu}}, \citenamefont
  {{Manning}}, \citenamefont {{McCormick}}, \citenamefont {{Meehan}},
  \citenamefont {{Randel}}, \citenamefont {{Rocken}}, \citenamefont
  {{Schreiner}}, \citenamefont {{Sokolovskiy}}, \citenamefont {{Syndergaard}},
  \citenamefont {{Thompson}}, \citenamefont {{Trenberth}}, \citenamefont
  {{Wee}}, \citenamefont {{Yen}},\ and\ \citenamefont
  {{Zeng}}}]{2008BAMS...89..313A}%
  \BibitemOpen
  \bibfield  {author} {\bibinfo {author} {\bibfnamefont {R.~A.}\ \bibnamefont
  {{Anthes}}}, \bibinfo {author} {\bibfnamefont {P.~A.}\ \bibnamefont
  {{Bernhardt}}}, \bibinfo {author} {\bibfnamefont {Y.}~\bibnamefont {{Chen}}},
  \bibinfo {author} {\bibfnamefont {L.}~\bibnamefont {{Cucurull}}}, \bibinfo
  {author} {\bibfnamefont {K.~F.}\ \bibnamefont {{Dymond}}}, \bibinfo {author}
  {\bibfnamefont {D.}~\bibnamefont {{Ector}}}, \bibinfo {author} {\bibfnamefont
  {S.~B.}\ \bibnamefont {{Healy}}}, \bibinfo {author} {\bibfnamefont {S.~P.}\
  \bibnamefont {{Ho}}}, \bibinfo {author} {\bibfnamefont {D.~C.}\ \bibnamefont
  {{Hunt}}}, \bibinfo {author} {\bibfnamefont {Y.~H.}\ \bibnamefont {{Kuo}}},
  \bibinfo {author} {\bibfnamefont {H.}~\bibnamefont {{Liu}}}, \bibinfo
  {author} {\bibfnamefont {K.}~\bibnamefont {{Manning}}}, \bibinfo {author}
  {\bibfnamefont {C.}~\bibnamefont {{McCormick}}}, \bibinfo {author}
  {\bibfnamefont {T.~K.}\ \bibnamefont {{Meehan}}}, \bibinfo {author}
  {\bibfnamefont {W.~J.}\ \bibnamefont {{Randel}}}, \bibinfo {author}
  {\bibfnamefont {C.}~\bibnamefont {{Rocken}}}, \bibinfo {author}
  {\bibfnamefont {W.~S.}\ \bibnamefont {{Schreiner}}}, \bibinfo {author}
  {\bibfnamefont {S.~V.}\ \bibnamefont {{Sokolovskiy}}}, \bibinfo {author}
  {\bibfnamefont {S.}~\bibnamefont {{Syndergaard}}}, \bibinfo {author}
  {\bibfnamefont {D.~C.}\ \bibnamefont {{Thompson}}}, \bibinfo {author}
  {\bibfnamefont {K.~E.}\ \bibnamefont {{Trenberth}}}, \bibinfo {author}
  {\bibfnamefont {T.~K.}\ \bibnamefont {{Wee}}}, \bibinfo {author}
  {\bibfnamefont {N.~L.}\ \bibnamefont {{Yen}}}, \ and\ \bibinfo {author}
  {\bibfnamefont {Z.}~\bibnamefont {{Zeng}}},\ }\href {\doibase
  10.1175/BAMS-89-3-313} {\bibfield  {journal} {\bibinfo  {journal} {Bulletin
  of the American Meteorological Society}\ }\textbf {\bibinfo {volume} {89}},\
  \bibinfo {pages} {313} (\bibinfo {year} {2008})}\BibitemShut {NoStop}%
\bibitem [{\citenamefont {{Fjeldbo}}\ \emph {et~al.}(1971)\citenamefont
  {{Fjeldbo}}, \citenamefont {{Kliore}},\ and\ \citenamefont
  {{Eshleman}}}]{1971AJ.....76..123F}%
  \BibitemOpen
  \bibfield  {author} {\bibinfo {author} {\bibfnamefont {G.}~\bibnamefont
  {{Fjeldbo}}}, \bibinfo {author} {\bibfnamefont {A.~J.}\ \bibnamefont
  {{Kliore}}}, \ and\ \bibinfo {author} {\bibfnamefont {V.~R.}\ \bibnamefont
  {{Eshleman}}},\ }\href {\doibase 10.1086/111096} {\bibfield  {journal}
  {\bibinfo  {journal} {\aj}\ }\textbf {\bibinfo {volume} {76}},\ \bibinfo
  {pages} {123} (\bibinfo {year} {1971})}\BibitemShut {NoStop}%
\bibitem [{\citenamefont {{Phinney}}\ and\ \citenamefont
  {{Anderson}}(1968)}]{1968JGR....73.1819P}%
  \BibitemOpen
  \bibfield  {author} {\bibinfo {author} {\bibfnamefont {R.~A.}\ \bibnamefont
  {{Phinney}}}\ and\ \bibinfo {author} {\bibfnamefont {D.~L.}\ \bibnamefont
  {{Anderson}}},\ }\href {\doibase 10.1029/JA073i005p01819} {\bibfield
  {journal} {\bibinfo  {journal} {\jgr}\ }\textbf {\bibinfo {volume} {73}},\
  \bibinfo {pages} {1819} (\bibinfo {year} {1968})}\BibitemShut {NoStop}%
\bibitem [{\citenamefont {{Lindal}}\ \emph {et~al.}(1981)\citenamefont
  {{Lindal}}, \citenamefont {{Wood}}, \citenamefont {{Levy}}, \citenamefont
  {{Anderson}}, \citenamefont {{Sweetnam}}, \citenamefont {{Hotz}},
  \citenamefont {{Buckles}}, \citenamefont {{Holmes}}, \citenamefont {{Doms}},
  \citenamefont {{Eshleman}}, \citenamefont {{Tyler}},\ and\ \citenamefont
  {{Croft}}}]{1981JGR....86.8721L}%
  \BibitemOpen
  \bibfield  {author} {\bibinfo {author} {\bibfnamefont {G.~F.}\ \bibnamefont
  {{Lindal}}}, \bibinfo {author} {\bibfnamefont {G.~E.}\ \bibnamefont
  {{Wood}}}, \bibinfo {author} {\bibfnamefont {G.~S.}\ \bibnamefont {{Levy}}},
  \bibinfo {author} {\bibfnamefont {J.~D.}\ \bibnamefont {{Anderson}}},
  \bibinfo {author} {\bibfnamefont {D.~N.}\ \bibnamefont {{Sweetnam}}},
  \bibinfo {author} {\bibfnamefont {H.~B.}\ \bibnamefont {{Hotz}}}, \bibinfo
  {author} {\bibfnamefont {B.~J.}\ \bibnamefont {{Buckles}}}, \bibinfo {author}
  {\bibfnamefont {D.~P.}\ \bibnamefont {{Holmes}}}, \bibinfo {author}
  {\bibfnamefont {P.~E.}\ \bibnamefont {{Doms}}}, \bibinfo {author}
  {\bibfnamefont {V.~R.}\ \bibnamefont {{Eshleman}}}, \bibinfo {author}
  {\bibfnamefont {G.~L.}\ \bibnamefont {{Tyler}}}, \ and\ \bibinfo {author}
  {\bibfnamefont {T.~A.}\ \bibnamefont {{Croft}}},\ }\href {\doibase
  10.1029/JA086iA10p08721} {\bibfield  {journal} {\bibinfo  {journal} {\jgr}\
  }\textbf {\bibinfo {volume} {86}},\ \bibinfo {pages} {8721} (\bibinfo {year}
  {1981})}\BibitemShut {NoStop}%
\bibitem [{\citenamefont {{Grandin}}\ \emph {et~al.}(2014)\citenamefont
  {{Grandin}}, \citenamefont {{Blelly}}, \citenamefont {{Witasse}},\ and\
  \citenamefont {{Marchaudon}}}]{2014JGRA..11910621G}%
  \BibitemOpen
  \bibfield  {author} {\bibinfo {author} {\bibfnamefont {M.}~\bibnamefont
  {{Grandin}}}, \bibinfo {author} {\bibfnamefont {P.~L.}\ \bibnamefont
  {{Blelly}}}, \bibinfo {author} {\bibfnamefont {O.}~\bibnamefont {{Witasse}}},
  \ and\ \bibinfo {author} {\bibfnamefont {A.}~\bibnamefont {{Marchaudon}}},\
  }\href {\doibase 10.1002/2014JA020698} {\bibfield  {journal} {\bibinfo
  {journal} {\jgr}\ }\textbf {\bibinfo {volume} {119}},\ \bibinfo {pages}
  {10,621} (\bibinfo {year} {2014})}\BibitemShut {NoStop}%
\bibitem [{\citenamefont {{Lipa}}\ and\ \citenamefont
  {{Tyler}}(1979)}]{1979Icar...39..192L}%
  \BibitemOpen
  \bibfield  {author} {\bibinfo {author} {\bibfnamefont {B.}~\bibnamefont
  {{Lipa}}}\ and\ \bibinfo {author} {\bibfnamefont {G.~L.}\ \bibnamefont
  {{Tyler}}},\ }\href {\doibase 10.1016/0019-1035(79)90163-5} {\bibfield
  {journal} {\bibinfo  {journal} {\icarus}\ }\textbf {\bibinfo {volume} {39}},\
  \bibinfo {pages} {192} (\bibinfo {year} {1979})}\BibitemShut {NoStop}%
\bibitem [{\citenamefont {{Jenkins}}\ \emph {et~al.}(1994)\citenamefont
  {{Jenkins}}, \citenamefont {{Steffes}}, \citenamefont {{Hinson}},
  \citenamefont {{Twicken}},\ and\ \citenamefont
  {{Tyler}}}]{1994Icar..110...79J}%
  \BibitemOpen
  \bibfield  {author} {\bibinfo {author} {\bibfnamefont {J.~M.}\ \bibnamefont
  {{Jenkins}}}, \bibinfo {author} {\bibfnamefont {P.~G.}\ \bibnamefont
  {{Steffes}}}, \bibinfo {author} {\bibfnamefont {D.~P.}\ \bibnamefont
  {{Hinson}}}, \bibinfo {author} {\bibfnamefont {J.~D.}\ \bibnamefont
  {{Twicken}}}, \ and\ \bibinfo {author} {\bibfnamefont {G.~L.}\ \bibnamefont
  {{Tyler}}},\ }\href {\doibase 10.1006/icar.1994.1108} {\bibfield  {journal}
  {\bibinfo  {journal} {\icarus}\ }\textbf {\bibinfo {volume} {110}},\ \bibinfo
  {pages} {79} (\bibinfo {year} {1994})}\BibitemShut {NoStop}%
\bibitem [{\citenamefont {{P{\"a}tzold}}\ \emph {et~al.}(2007)\citenamefont
  {{P{\"a}tzold}}, \citenamefont {{H{\"a}usler}}, \citenamefont {{Bird}},
  \citenamefont {{Tellmann}}, \citenamefont {{Mattei}}, \citenamefont
  {{Asmar}}, \citenamefont {{Dehant}}, \citenamefont {{Eidel}}, \citenamefont
  {{Imamura}}, \citenamefont {{Simpson}},\ and\ \citenamefont
  {{Tyler}}}]{2007Natur.450..657P}%
  \BibitemOpen
  \bibfield  {author} {\bibinfo {author} {\bibfnamefont {M.}~\bibnamefont
  {{P{\"a}tzold}}}, \bibinfo {author} {\bibfnamefont {B.}~\bibnamefont
  {{H{\"a}usler}}}, \bibinfo {author} {\bibfnamefont {M.~K.}\ \bibnamefont
  {{Bird}}}, \bibinfo {author} {\bibfnamefont {S.}~\bibnamefont {{Tellmann}}},
  \bibinfo {author} {\bibfnamefont {R.}~\bibnamefont {{Mattei}}}, \bibinfo
  {author} {\bibfnamefont {S.~W.}\ \bibnamefont {{Asmar}}}, \bibinfo {author}
  {\bibfnamefont {V.}~\bibnamefont {{Dehant}}}, \bibinfo {author}
  {\bibfnamefont {W.}~\bibnamefont {{Eidel}}}, \bibinfo {author} {\bibfnamefont
  {T.}~\bibnamefont {{Imamura}}}, \bibinfo {author} {\bibfnamefont {R.~A.}\
  \bibnamefont {{Simpson}}}, \ and\ \bibinfo {author} {\bibfnamefont {G.~L.}\
  \bibnamefont {{Tyler}}},\ }\href {\doibase 10.1038/nature06239} {\bibfield
  {journal} {\bibinfo  {journal} {\nat}\ }\textbf {\bibinfo {volume} {450}},\
  \bibinfo {pages} {657} (\bibinfo {year} {2007})}\BibitemShut {NoStop}%
\bibitem [{\citenamefont {{Tellmann}}\ \emph {et~al.}(2009)\citenamefont
  {{Tellmann}}, \citenamefont {{P{\"a}tzold}}, \citenamefont {{H{\"a}usler}},
  \citenamefont {{Bird}},\ and\ \citenamefont {{Tyler}}}]{2009JGRE..114.0B36T}%
  \BibitemOpen
  \bibfield  {author} {\bibinfo {author} {\bibfnamefont {S.}~\bibnamefont
  {{Tellmann}}}, \bibinfo {author} {\bibfnamefont {M.}~\bibnamefont
  {{P{\"a}tzold}}}, \bibinfo {author} {\bibfnamefont {B.}~\bibnamefont
  {{H{\"a}usler}}}, \bibinfo {author} {\bibfnamefont {M.~K.}\ \bibnamefont
  {{Bird}}}, \ and\ \bibinfo {author} {\bibfnamefont {G.~L.}\ \bibnamefont
  {{Tyler}}},\ }\href {\doibase 10.1029/2008JE003204} {\bibfield  {journal}
  {\bibinfo  {journal} {\jgr}\ }\textbf {\bibinfo {volume} {114}},\ \bibinfo
  {eid} {E00B36} (\bibinfo {year} {2009})}\BibitemShut {NoStop}%
\bibitem [{\citenamefont {{Hinson}}\ \emph {et~al.}(1999)\citenamefont
  {{Hinson}}, \citenamefont {{Simpson}}, \citenamefont {{Twicken}},
  \citenamefont {{Tyler}},\ and\ \citenamefont
  {{Flasar}}}]{1999JGR...10426997H}%
  \BibitemOpen
  \bibfield  {author} {\bibinfo {author} {\bibfnamefont {D.~P.}\ \bibnamefont
  {{Hinson}}}, \bibinfo {author} {\bibfnamefont {R.~A.}\ \bibnamefont
  {{Simpson}}}, \bibinfo {author} {\bibfnamefont {J.~D.}\ \bibnamefont
  {{Twicken}}}, \bibinfo {author} {\bibfnamefont {G.~L.}\ \bibnamefont
  {{Tyler}}}, \ and\ \bibinfo {author} {\bibfnamefont {F.~M.}\ \bibnamefont
  {{Flasar}}},\ }\href {\doibase 10.1029/1999JE001069} {\bibfield  {journal}
  {\bibinfo  {journal} {\jgr}\ }\textbf {\bibinfo {volume} {104}},\ \bibinfo
  {pages} {26997} (\bibinfo {year} {1999})}\BibitemShut {NoStop}%
\bibitem [{\citenamefont {{P{\"a}tzold}}\ \emph {et~al.}(2016)\citenamefont
  {{P{\"a}tzold}}, \citenamefont {{H{\"a}usler}}, \citenamefont {{Tyler}},
  \citenamefont {{Andert}}, \citenamefont {{Asmar}}, \citenamefont {{Bird}},
  \citenamefont {{Dehant}}, \citenamefont {{Hinson}}, \citenamefont
  {{Rosenblatt}}, \citenamefont {{Simpson}}, \citenamefont {{Tellmann}},
  \citenamefont {{Withers}}, \citenamefont {{Beuthe}}, \citenamefont
  {{Efimov}}, \citenamefont {{Hahn}}, \citenamefont {{Kahan}}, \citenamefont
  {{Le Maistre}}, \citenamefont {{Oschlisniok}}, \citenamefont {{Peter}},\ and\
  \citenamefont {{Remus}}}]{2016P&SS..127...44P}%
  \BibitemOpen
  \bibfield  {author} {\bibinfo {author} {\bibfnamefont {M.}~\bibnamefont
  {{P{\"a}tzold}}}, \bibinfo {author} {\bibfnamefont {B.}~\bibnamefont
  {{H{\"a}usler}}}, \bibinfo {author} {\bibfnamefont {G.~L.}\ \bibnamefont
  {{Tyler}}}, \bibinfo {author} {\bibfnamefont {T.}~\bibnamefont {{Andert}}},
  \bibinfo {author} {\bibfnamefont {S.~W.}\ \bibnamefont {{Asmar}}}, \bibinfo
  {author} {\bibfnamefont {M.~K.}\ \bibnamefont {{Bird}}}, \bibinfo {author}
  {\bibfnamefont {V.}~\bibnamefont {{Dehant}}}, \bibinfo {author}
  {\bibfnamefont {D.~P.}\ \bibnamefont {{Hinson}}}, \bibinfo {author}
  {\bibfnamefont {P.}~\bibnamefont {{Rosenblatt}}}, \bibinfo {author}
  {\bibfnamefont {R.~A.}\ \bibnamefont {{Simpson}}}, \bibinfo {author}
  {\bibfnamefont {S.}~\bibnamefont {{Tellmann}}}, \bibinfo {author}
  {\bibfnamefont {P.}~\bibnamefont {{Withers}}}, \bibinfo {author}
  {\bibfnamefont {M.}~\bibnamefont {{Beuthe}}}, \bibinfo {author}
  {\bibfnamefont {A.~I.}\ \bibnamefont {{Efimov}}}, \bibinfo {author}
  {\bibfnamefont {M.}~\bibnamefont {{Hahn}}}, \bibinfo {author} {\bibfnamefont
  {D.}~\bibnamefont {{Kahan}}}, \bibinfo {author} {\bibfnamefont
  {S.}~\bibnamefont {{Le Maistre}}}, \bibinfo {author} {\bibfnamefont
  {J.}~\bibnamefont {{Oschlisniok}}}, \bibinfo {author} {\bibfnamefont
  {K.}~\bibnamefont {{Peter}}}, \ and\ \bibinfo {author} {\bibfnamefont
  {S.}~\bibnamefont {{Remus}}},\ }\href {\doibase 10.1016/j.pss.2016.02.013}
  {\bibfield  {journal} {\bibinfo  {journal} {\planss}\ }\textbf {\bibinfo
  {volume} {127}},\ \bibinfo {pages} {44} (\bibinfo {year} {2016})}\BibitemShut
  {NoStop}%
\bibitem [{\citenamefont {{Schinder}}\ \emph {et~al.}(2011)\citenamefont
  {{Schinder}}, \citenamefont {{Flasar}}, \citenamefont {{Marouf}},
  \citenamefont {{French}}, \citenamefont {{McGhee}}, \citenamefont {{Kliore}},
  \citenamefont {{Rappaport}}, \citenamefont {{Barbinis}}, \citenamefont
  {{Fleischman}},\ and\ \citenamefont {{Anabtawi}}}]{2011Icar..215..460S}%
  \BibitemOpen
  \bibfield  {author} {\bibinfo {author} {\bibfnamefont {P.~J.}\ \bibnamefont
  {{Schinder}}}, \bibinfo {author} {\bibfnamefont {F.~M.}\ \bibnamefont
  {{Flasar}}}, \bibinfo {author} {\bibfnamefont {E.~A.}\ \bibnamefont
  {{Marouf}}}, \bibinfo {author} {\bibfnamefont {R.~G.}\ \bibnamefont
  {{French}}}, \bibinfo {author} {\bibfnamefont {C.~A.}\ \bibnamefont
  {{McGhee}}}, \bibinfo {author} {\bibfnamefont {A.~J.}\ \bibnamefont
  {{Kliore}}}, \bibinfo {author} {\bibfnamefont {N.~J.}\ \bibnamefont
  {{Rappaport}}}, \bibinfo {author} {\bibfnamefont {E.}~\bibnamefont
  {{Barbinis}}}, \bibinfo {author} {\bibfnamefont {D.}~\bibnamefont
  {{Fleischman}}}, \ and\ \bibinfo {author} {\bibfnamefont {A.}~\bibnamefont
  {{Anabtawi}}},\ }\href {\doibase 10.1016/j.icarus.2011.07.030} {\bibfield
  {journal} {\bibinfo  {journal} {\icarus}\ }\textbf {\bibinfo {volume}
  {215}},\ \bibinfo {pages} {460} (\bibinfo {year} {2011})}\BibitemShut
  {NoStop}%
\bibitem [{\citenamefont {{Petricca}}\ \emph {et~al.}(2021)\citenamefont
  {{Petricca}}, \citenamefont {{Cascioli}},\ and\ \citenamefont
  {{Genova}}}]{2021RaSc...5607205P}%
  \BibitemOpen
  \bibfield  {author} {\bibinfo {author} {\bibfnamefont {F.}~\bibnamefont
  {{Petricca}}}, \bibinfo {author} {\bibfnamefont {G.}~\bibnamefont
  {{Cascioli}}}, \ and\ \bibinfo {author} {\bibfnamefont {A.}~\bibnamefont
  {{Genova}}},\ }\href {\doibase 10.1029/2020RS007205} {\bibfield  {journal}
  {\bibinfo  {journal} {Radio Science}\ }\textbf {\bibinfo {volume} {56}},\
  \bibinfo {eid} {e07205} (\bibinfo {year} {2021})}\BibitemShut {NoStop}%
\bibitem [{\citenamefont {{Withers}}(2010)}]{2010AdSpR..46...58W}%
  \BibitemOpen
  \bibfield  {author} {\bibinfo {author} {\bibfnamefont {P.}~\bibnamefont
  {{Withers}}},\ }\href {\doibase 10.1016/j.asr.2010.03.004} {\bibfield
  {journal} {\bibinfo  {journal} {\asr}\ }\textbf {\bibinfo {volume} {46}},\
  \bibinfo {pages} {58} (\bibinfo {year} {2010})}\BibitemShut {NoStop}%
\bibitem [{\citenamefont {{Withers}}(2020)}]{2020AdSpR..66.2466W}%
  \BibitemOpen
  \bibfield  {author} {\bibinfo {author} {\bibfnamefont {P.}~\bibnamefont
  {{Withers}}},\ }\href {\doibase 10.1016/j.asr.2020.07.049} {\bibfield
  {journal} {\bibinfo  {journal} {\asr}\ }\textbf {\bibinfo {volume} {66}},\
  \bibinfo {pages} {2466} (\bibinfo {year} {2020})}\BibitemShut {NoStop}%
\bibitem [{\citenamefont {Bourgoin}(2020)}]{PhysRevD.101.064035}%
  \BibitemOpen
  \bibfield  {author} {\bibinfo {author} {\bibfnamefont {A.}~\bibnamefont
  {Bourgoin}},\ }\href {\doibase 10.1103/PhysRevD.101.064035} {\bibfield
  {journal} {\bibinfo  {journal} {\prd}\ }\textbf {\bibinfo {volume} {101}},\
  \bibinfo {pages} {064035} (\bibinfo {year} {2020})}\BibitemShut {NoStop}%
\bibitem [{\citenamefont {Linet}\ and\ \citenamefont
  {Teyssandier}(2002)}]{2002PhRvD..66b4045L}%
  \BibitemOpen
  \bibfield  {author} {\bibinfo {author} {\bibfnamefont {B.}~\bibnamefont
  {Linet}}\ and\ \bibinfo {author} {\bibfnamefont {P.}~\bibnamefont
  {Teyssandier}},\ }\href {\doibase 10.1103/PhysRevD.66.024045} {\bibfield
  {journal} {\bibinfo  {journal} {\prd}\ }\textbf {\bibinfo {volume} {66}},\
  \bibinfo {pages} {024045} (\bibinfo {year} {2002})}\BibitemShut {NoStop}%
\bibitem [{\citenamefont {Le~Poncin-Lafitte}\ \emph {et~al.}(2004)\citenamefont
  {Le~Poncin-Lafitte}, \citenamefont {Linet},\ and\ \citenamefont
  {Teyssandier}}]{2004CQGra..21.4463L}%
  \BibitemOpen
  \bibfield  {author} {\bibinfo {author} {\bibfnamefont {C.}~\bibnamefont
  {Le~Poncin-Lafitte}}, \bibinfo {author} {\bibfnamefont {B.}~\bibnamefont
  {Linet}}, \ and\ \bibinfo {author} {\bibfnamefont {P.}~\bibnamefont
  {Teyssandier}},\ }\href {\doibase 10.1088/0264-9381/21/18/012} {\bibfield
  {journal} {\bibinfo  {journal} {\cqg}\ }\textbf {\bibinfo {volume} {21}},\
  \bibinfo {pages} {4463} (\bibinfo {year} {2004})}\BibitemShut {NoStop}%
\bibitem [{\citenamefont {Gordon}(1923)}]{doi101002andp19233772202}%
  \BibitemOpen
  \bibfield  {author} {\bibinfo {author} {\bibfnamefont {W.}~\bibnamefont
  {Gordon}},\ }\href {\doibase 10.1002/andp.19233772202} {\bibfield  {journal}
  {\bibinfo  {journal} {Annalen der Physik}\ }\textbf {\bibinfo {volume}
  {377}},\ \bibinfo {pages} {421} (\bibinfo {year} {1923})}\BibitemShut
  {NoStop}%
\bibitem [{\citenamefont {{Bourgoin}}\ \emph {et~al.}(2021)\citenamefont
  {{Bourgoin}}, \citenamefont {{Zannoni}}, \citenamefont {{Gomez Casajus}},
  \citenamefont {{Tortora}},\ and\ \citenamefont
  {{Teyssandier}}}]{2021A&A...648A..46B}%
  \BibitemOpen
  \bibfield  {author} {\bibinfo {author} {\bibfnamefont {A.}~\bibnamefont
  {{Bourgoin}}}, \bibinfo {author} {\bibfnamefont {M.}~\bibnamefont
  {{Zannoni}}}, \bibinfo {author} {\bibfnamefont {L.}~\bibnamefont {{Gomez
  Casajus}}}, \bibinfo {author} {\bibfnamefont {P.}~\bibnamefont {{Tortora}}},
  \ and\ \bibinfo {author} {\bibfnamefont {P.}~\bibnamefont {{Teyssandier}}},\
  }\href {\doibase 10.1051/0004-6361/202040269} {\bibfield  {journal} {\bibinfo
   {journal} {\aap}\ }\textbf {\bibinfo {volume} {648}},\ \bibinfo {eid} {A46}
  (\bibinfo {year} {2021})}\BibitemShut {NoStop}%
\bibitem [{\citenamefont {{Bourgoin}}\ \emph {et~al.}(2019)\citenamefont
  {{Bourgoin}}, \citenamefont {{Zannoni}},\ and\ \citenamefont
  {{Tortora}}}]{2019A&A...624A..41B}%
  \BibitemOpen
  \bibfield  {author} {\bibinfo {author} {\bibfnamefont {A.}~\bibnamefont
  {{Bourgoin}}}, \bibinfo {author} {\bibfnamefont {M.}~\bibnamefont
  {{Zannoni}}}, \ and\ \bibinfo {author} {\bibfnamefont {P.}~\bibnamefont
  {{Tortora}}},\ }\href {\doibase 10.1051/0004-6361/201834962} {\bibfield
  {journal} {\bibinfo  {journal} {\aap}\ }\textbf {\bibinfo {volume} {624}},\
  \bibinfo {eid} {A41} (\bibinfo {year} {2019})}\BibitemShut {NoStop}%
\bibitem [{\citenamefont {Synge}(1960)}]{SyngeBookGR}%
  \BibitemOpen
  \bibfield  {author} {\bibinfo {author} {\bibfnamefont {J.~L.}\ \bibnamefont
  {Synge}},\ }\href {https://ui.adsabs.harvard.edu/abs/1960rgt..book.....S}
  {\emph {\bibinfo {title} {Relativity: The General Theory}}}\ (\bibinfo
  {publisher} {North-Holland Publ. Co.},\ \bibinfo {address} {Amsterdam},\
  \bibinfo {year} {1960})\BibitemShut {NoStop}%
\bibitem [{\citenamefont {Blanchet}\ \emph {et~al.}(2001)\citenamefont
  {Blanchet}, \citenamefont {Salomon}, \citenamefont {Teyssandier},\ and\
  \citenamefont {Wolf}}]{2001A&A...370..320B}%
  \BibitemOpen
  \bibfield  {author} {\bibinfo {author} {\bibfnamefont {L.}~\bibnamefont
  {Blanchet}}, \bibinfo {author} {\bibfnamefont {C.}~\bibnamefont {Salomon}},
  \bibinfo {author} {\bibfnamefont {P.}~\bibnamefont {Teyssandier}}, \ and\
  \bibinfo {author} {\bibfnamefont {P.}~\bibnamefont {Wolf}},\ }\href {\doibase
  10.1051/0004-6361:20010233} {\bibfield  {journal} {\bibinfo  {journal}
  {\aap}\ }\textbf {\bibinfo {volume} {370}},\ \bibinfo {pages} {320} (\bibinfo
  {year} {2001})}\BibitemShut {NoStop}%
\bibitem [{\citenamefont {Hees}\ \emph {et~al.}(2012)\citenamefont {Hees},
  \citenamefont {Lamine}, \citenamefont {Reynaud}, \citenamefont {{Jaekel}},
  \citenamefont {{Le Poncin-Lafitte}}, \citenamefont {{Lainey}}, \citenamefont
  {{F{\"u}zfa}}, \citenamefont {{Courty}}, \citenamefont {{Dehant}},\ and\
  \citenamefont {{Wolf}}}]{2012CQGra..29w5027H}%
  \BibitemOpen
  \bibfield  {author} {\bibinfo {author} {\bibfnamefont {A.}~\bibnamefont
  {Hees}}, \bibinfo {author} {\bibfnamefont {B.}~\bibnamefont {Lamine}},
  \bibinfo {author} {\bibfnamefont {S.}~\bibnamefont {Reynaud}}, \bibinfo
  {author} {\bibfnamefont {M.-T.}\ \bibnamefont {{Jaekel}}}, \bibinfo {author}
  {\bibfnamefont {C.}~\bibnamefont {{Le Poncin-Lafitte}}}, \bibinfo {author}
  {\bibfnamefont {V.}~\bibnamefont {{Lainey}}}, \bibinfo {author}
  {\bibfnamefont {A.}~\bibnamefont {{F{\"u}zfa}}}, \bibinfo {author}
  {\bibfnamefont {J.-M.}\ \bibnamefont {{Courty}}}, \bibinfo {author}
  {\bibfnamefont {V.}~\bibnamefont {{Dehant}}}, \ and\ \bibinfo {author}
  {\bibfnamefont {P.}~\bibnamefont {{Wolf}}},\ }\href {\doibase
  10.1088/0264-9381/29/23/235027} {\bibfield  {journal} {\bibinfo  {journal}
  {\cqg}\ }\textbf {\bibinfo {volume} {29}},\ \bibinfo {pages} {235027}
  (\bibinfo {year} {2012})}\BibitemShut {NoStop}%
\bibitem [{\citenamefont {{Hees}}\ \emph {et~al.}(2014)\citenamefont {{Hees}},
  \citenamefont {{Bertone}},\ and\ \citenamefont {{Le
  Poncin-Lafitte}}}]{2014PhRvD..89f4045H}%
  \BibitemOpen
  \bibfield  {author} {\bibinfo {author} {\bibfnamefont {A.}~\bibnamefont
  {{Hees}}}, \bibinfo {author} {\bibfnamefont {S.}~\bibnamefont {{Bertone}}}, \
  and\ \bibinfo {author} {\bibfnamefont {C.}~\bibnamefont {{Le
  Poncin-Lafitte}}},\ }\href {\doibase 10.1103/PhysRevD.89.064045} {\bibfield
  {journal} {\bibinfo  {journal} {\prd}\ }\textbf {\bibinfo {volume} {89}},\
  \bibinfo {eid} {064045} (\bibinfo {year} {2014})}\BibitemShut {NoStop}%
\bibitem [{\citenamefont {Teyssandier}\ and\ \citenamefont
  {Le~Poncin-Lafitte}(2008)}]{2008CQGra25n5020T}%
  \BibitemOpen
  \bibfield  {author} {\bibinfo {author} {\bibfnamefont {P.}~\bibnamefont
  {Teyssandier}}\ and\ \bibinfo {author} {\bibfnamefont {C.}~\bibnamefont
  {Le~Poncin-Lafitte}},\ }\href {\doibase 10.1088/0264-9381/25/14/145020}
  {\bibfield  {journal} {\bibinfo  {journal} {\cqg}\ }\textbf {\bibinfo
  {volume} {25}},\ \bibinfo {pages} {145020} (\bibinfo {year}
  {2008})}\BibitemShut {NoStop}%
\bibitem [{\citenamefont {Linet}\ and\ \citenamefont
  {Teyssandier}(2016)}]{PhysRevD.93.044028}%
  \BibitemOpen
  \bibfield  {author} {\bibinfo {author} {\bibfnamefont {B.}~\bibnamefont
  {Linet}}\ and\ \bibinfo {author} {\bibfnamefont {P.}~\bibnamefont
  {Teyssandier}},\ }\href {\doibase 10.1103/PhysRevD.93.044028} {\bibfield
  {journal} {\bibinfo  {journal} {\prd}\ }\textbf {\bibinfo {volume} {93}},\
  \bibinfo {pages} {044028} (\bibinfo {year} {2016})}\BibitemShut {NoStop}%
\bibitem [{\citenamefont {Arfken}(1985)}]{garfken67math}%
  \BibitemOpen
  \bibfield  {author} {\bibinfo {author} {\bibfnamefont {G.}~\bibnamefont
  {Arfken}},\ }\href@noop {} {\emph {\bibinfo {title} {Mathematical Methods for
  Physicists}}},\ \bibinfo {edition} {3rd}\ ed.\ (\bibinfo  {publisher}
  {Academic Press, {Inc.}},\ \bibinfo {address} {San Diego},\ \bibinfo {year}
  {1985})\BibitemShut {NoStop}%
\bibitem [{\citenamefont {{Withers}}\ \emph {et~al.}(2014)\citenamefont
  {{Withers}}, \citenamefont {{Moore}}, \citenamefont {{Cahoy}},\ and\
  \citenamefont {{Beerer}}}]{2014P&SS..101...77W}%
  \BibitemOpen
  \bibfield  {author} {\bibinfo {author} {\bibfnamefont {P.}~\bibnamefont
  {{Withers}}}, \bibinfo {author} {\bibfnamefont {L.}~\bibnamefont {{Moore}}},
  \bibinfo {author} {\bibfnamefont {K.}~\bibnamefont {{Cahoy}}}, \ and\
  \bibinfo {author} {\bibfnamefont {I.}~\bibnamefont {{Beerer}}},\ }\href
  {\doibase 10.1016/j.pss.2014.06.011} {\bibfield  {journal} {\bibinfo
  {journal} {\planss}\ }\textbf {\bibinfo {volume} {101}},\ \bibinfo {pages}
  {77} (\bibinfo {year} {2014})}\BibitemShut {NoStop}%
\bibitem [{\citenamefont {{Withers}}\ and\ \citenamefont
  {{Moore}}(2020)}]{2020RaSc...5507046W}%
  \BibitemOpen
  \bibfield  {author} {\bibinfo {author} {\bibfnamefont {P.}~\bibnamefont
  {{Withers}}}\ and\ \bibinfo {author} {\bibfnamefont {L.}~\bibnamefont
  {{Moore}}},\ }\href {\doibase 10.1029/2019RS007046} {\bibfield  {journal}
  {\bibinfo  {journal} {Radio Science}\ }\textbf {\bibinfo {volume} {55}},\
  \bibinfo {eid} {e07046} (\bibinfo {year} {2020})}\BibitemShut {NoStop}%
\bibitem [{\citenamefont {{Eshleman}}(1973)}]{1973P&SS...21.1521E}%
  \BibitemOpen
  \bibfield  {author} {\bibinfo {author} {\bibfnamefont {V.~R.}\ \bibnamefont
  {{Eshleman}}},\ }\href {\doibase 10.1016/0032-0633(73)90059-7} {\bibfield
  {journal} {\bibinfo  {journal} {\planss}\ }\textbf {\bibinfo {volume} {21}},\
  \bibinfo {pages} {1521} (\bibinfo {year} {1973})}\BibitemShut {NoStop}%
\bibitem [{\citenamefont {{Acton}}(1996)}]{1996P&SS...44...65Aq}%
  \BibitemOpen
  \bibfield  {author} {\bibinfo {author} {\bibfnamefont {C.~H.}\ \bibnamefont
  {{Acton}}},\ }\href {\doibase 10.1016/0032-0633(95)00107-7} {\bibfield
  {journal} {\bibinfo  {journal} {\planss}\ }\textbf {\bibinfo {volume} {44}},\
  \bibinfo {pages} {65} (\bibinfo {year} {1996})}\BibitemShut {NoStop}%
\bibitem [{\citenamefont {{Bender}}(1973)}]{1973Cryo...13...11B}%
  \BibitemOpen
  \bibfield  {author} {\bibinfo {author} {\bibfnamefont {E.}~\bibnamefont
  {{Bender}}},\ }\href {\doibase 10.1016/0011-2275(73)90258-0} {\bibfield
  {journal} {\bibinfo  {journal} {Cryogenics}\ }\textbf {\bibinfo {volume}
  {13}},\ \bibinfo {pages} {11} (\bibinfo {year} {1973})}\BibitemShut {NoStop}%
\bibitem [{\citenamefont {{B{\"u}hner}}\ \emph {et~al.}(1981)\citenamefont
  {{B{\"u}hner}}, \citenamefont {{Maurer}},\ and\ \citenamefont
  {{Bender}}}]{1981Cryo...21..157B}%
  \BibitemOpen
  \bibfield  {author} {\bibinfo {author} {\bibfnamefont {K.}~\bibnamefont
  {{B{\"u}hner}}}, \bibinfo {author} {\bibfnamefont {G.}~\bibnamefont
  {{Maurer}}}, \ and\ \bibinfo {author} {\bibfnamefont {E.}~\bibnamefont
  {{Bender}}},\ }\href {\doibase 10.1016/0011-2275(81)90267-8} {\bibfield
  {journal} {\bibinfo  {journal} {Cryogenics}\ }\textbf {\bibinfo {volume}
  {21}},\ \bibinfo {pages} {157} (\bibinfo {year} {1981})}\BibitemShut
  {NoStop}%
\bibitem [{\citenamefont {{Gramigna}}\ \emph {et~al.}(2021)\citenamefont
  {{Gramigna}}, \citenamefont {{Parisi}}, \citenamefont {{Buccino}},
  \citenamefont {{Gomez Casajus}}, \citenamefont {{Zannoni}}, \citenamefont
  {{Tortora}},\ and\ \citenamefont {{Oudrhiri}}}]{2021arXiv210208300G}%
  \BibitemOpen
  \bibfield  {author} {\bibinfo {author} {\bibfnamefont {E.}~\bibnamefont
  {{Gramigna}}}, \bibinfo {author} {\bibfnamefont {M.}~\bibnamefont
  {{Parisi}}}, \bibinfo {author} {\bibfnamefont {D.}~\bibnamefont {{Buccino}}},
  \bibinfo {author} {\bibfnamefont {L.}~\bibnamefont {{Gomez Casajus}}},
  \bibinfo {author} {\bibfnamefont {M.}~\bibnamefont {{Zannoni}}}, \bibinfo
  {author} {\bibfnamefont {P.}~\bibnamefont {{Tortora}}}, \ and\ \bibinfo
  {author} {\bibfnamefont {K.}~\bibnamefont {{Oudrhiri}}},\ }\href@noop {}
  {\bibfield  {journal} {\bibinfo  {journal} {arXiv e-prints}\ } (\bibinfo
  {year} {2021})},\ \Eprint {http://arxiv.org/abs/2102.08300} {arXiv:2102.08300
  [astro-ph.EP]} \BibitemShut {NoStop}%
\bibitem [{\citenamefont {{Withers}}\ \emph {et~al.}(2003)\citenamefont
  {{Withers}}, \citenamefont {{Towner}}, \citenamefont {{Hathi}},\ and\
  \citenamefont {{Zarnecki}}}]{2003P&SS...51..541W}%
  \BibitemOpen
  \bibfield  {author} {\bibinfo {author} {\bibfnamefont {P.}~\bibnamefont
  {{Withers}}}, \bibinfo {author} {\bibfnamefont {M.~C.}\ \bibnamefont
  {{Towner}}}, \bibinfo {author} {\bibfnamefont {B.}~\bibnamefont {{Hathi}}}, \
  and\ \bibinfo {author} {\bibfnamefont {J.~C.}\ \bibnamefont {{Zarnecki}}},\
  }\href {\doibase 10.1016/S0032-0633(03)00077-1} {\bibfield  {journal}
  {\bibinfo  {journal} {\planss}\ }\textbf {\bibinfo {volume} {51}},\ \bibinfo
  {pages} {541} (\bibinfo {year} {2003})}\BibitemShut {NoStop}%
\end{thebibliography}%

\end{document}